\documentclass[preprint2,times,tighten]{aastex62}

\shorttitle{X-ray spectral analysis of stars in the Cyg\,OB2 region}
\shortauthors{Flaccomio et al.}

\usepackage{natbib}
\begin{document}


\title{X-ray spectral characterization of the young Cygnus OB2 population}

\author{E. Flaccomio}
\affiliation{INAF-Osservatorio Astronomico di Palermo Giuseppe S. Vaiana, Piazza del Parlamento 1, 90134 Palermo, Italy}
\author{J. F. Albacete-Colombo}
\affiliation{Universidad de Rio Negro, Sede Atl\'antica - CONICET, Viedma CP8500, Argentina}
\author{J. J. Drake}
\affiliation{Smithsonian Astrophysical Observatory, 60 Garden St., Cambridge, MA 02138, U.S.A}
\author{M. G. Guarcello}
\affiliation{INAF-Osservatorio Astronomico di Palermo Giuseppe S. Vaiana, Piazza del Parlamento 1, 90134 Palermo, Italy}
\author{V. Kashyap}
\affiliation{Smithsonian Astrophysical Observatory, 60 Garden St., Cambridge, MA 02138, U.S.A}
\author{N. J. Wright}
\affiliation{Astrophysics Group, Keele University, Keele, ST5 5BG, UK}
\author{K. Briggs}
\affiliation{Hamburger Sternwarte, Universität Hamburg, Gojenbergsweg 112, 21029, Hamburg, Germany
}
\affiliation{Smithsonian Astrophysical Observatory, 60 Garden St., Cambridge, MA 02138, U.S.A}
\author{B. Ercolano}
\affiliation{Universit\"ats-Sternwarte M\"unchen, Scheinerstr. 1, D-81679 M\"unchen, Germany}
\affiliation{Excellence Cluster Origin and Structure of the Universe, Boltzmannstr. 2, D-85748 Garching bei M\"unchen, Germany}
\author{M. McCollogh}
\affiliation{Smithsonian Astrophysical Observatory, 60 Garden St., Cambridge, MA 02138, U.S.A}
\author{S. Sciortino}\affiliation{INAF-Osservatorio Astronomico di Palermo Giuseppe S. Vaiana, Piazza del Parlamento 1, 90134 Palermo, Italy}

\email{ettoref@astropa.inaf.it}


\begin{abstract}

We analyze the X-ray spectra of the $\sim$8000 sources detected in the
Cygnus\,OB2 {\em Chandra} Legacy Survey (Drake et al., this issue), with
the goals of characterizing the coronal plasma of the young
low-mass stars in the region and estimating their intrinsic X-ray luminosities. We adopt two different strategies for
X-ray sources for which more or less than 20  photons were
detected. For the brighter sample we fit the spectra with absorbed
isothermal models. In order to limit uncertainties, for most of the
fainter Cygnus\,OB2 members in this sample, we constrain the spectral
parameters to characteristic ranges defined from the brightest stars.
For X-ray sources with $<$20 net photons we adopt a conversion factor
from detected photon flux to intrinsic flux. This was defined, building
on the results for the previous sample, as a function of the
20\% quantile of the detected photon energy distributions, which we
prove to also correlate well with extinction. We then use the X-ray extinction from the spectral fits to constrain the ratio between optical and X-ray extinction toward Cygnus\,OB2, finding it consistent with standard ``Galactic'' values, when properly accounting for systematics. Finally we exploit the large number of sources to constrain the average coronal abundances of several elements, through two different ensemble analyses of the X-ray spectra of low-mass Cygnus\,OB2 members. We find the pattern of abundances to be largely consistent with that derived for the young stellar coronae in the Orion Nebula Cluster.

\end{abstract}



\section{Introduction}

The Cygnus\,OB2 association is one of the most massive groups of young stars in the galaxy, harboring hundreds of OB 
stars \citep[e.g.,][]{sch56a,mas91a,com02a,han03a,kim07a,wri15a} and tens of thousands 
of lower-mass, pre-main sequence stars \citep[e.g.,][]{alb07,dre08a,vin08a,wri09}. 
Embedded within the wider Cygnus\,X giant molecular cloud \citep{sch06}, the 
association is the dominant source of feedback for the region \citep{wri12a,gua16a}. 
Its size and proximity make Cyg\,OB2 the best available environment to study the formation 
and evolution of large OB associations \citep{wri14b,wri16a}, the properties of 
both low- and high-mass stars \citep[e.g.,][]{kim12a,rau15}, and the evolution
of protoplanetary disks \citep{gua13a}.

The {\em Chandra} Cygnus OB2 Legacy Survey (Drake et al., this issue) is 
designed to explore many of these issues by providing a wide-area and uniform
census of both the high-mass and low-mass stars in Cyg\,OB2 and their X-ray properties. 
The survey is composed of a mosaic of overlapping {\em Chandra} ACIS-I pointings covering 
the central 1 degree area of the massive young stellar association
Cygnus\,OB2. The observations and resulting source catalog are described 
in \citet{wri14a}, with an analysis of the sensitivity and completeness of 
the observations in \citet{wri15b}. \citet{gua15a} correlates the X-ray catalog 
with available optical and infrared (IR) photometric catalogs together with 
new deep optical photometry. \citet{kas18} 
exploits the X-ray and
optical/IR information to assess the relation of the X-ray sources with
the Cygnus\,OB2 region. In short, $\sim$8000 X-ray point sources were
detected, $\sim$6000 of which are associated with young members of the
Cygnus\,OB2 association.  The remainder are associated with interloping field stars, mostly
in the foreground, and with background extragalactic sources, mostly active
galactic nuclei (AGNs). 

Scientific motivation for studying the X-ray properties of Cygnus\,OB2 stars falls into two broad categories.  Firstly, the association represents a slightly older group of pre-main sequence stars than has been studied in detail in similar large surveys, such as the Orion Nebula Cluster (ONC; \citealt{hil97,get05}).  Stars in the ONC have been estimated to have an average age of about 2-3 Myr \citep{dar10}, whereas \citet{wri10a} and \citet{wri15a} have found Cygnus~OB2 to have a mean age of about 5 Myr, but with an age spread of about 3-4 Myr. Stars in Cygnus~OB2 therefore represent an important large sample with which to assess how magnetic activity and X-ray emission evolves in the first few million years that are key to protoplanetary disk evolution and planet formation.

Secondly, basic X-ray spectral information is also used by
\citet{kas18} for their membership analysis, exploiting the fact
that the X-ray spectra of members, foreground, and background
sources differ significantly, at least statistically. The strong and hard 
X-ray emission of PMS stars with respect to
that of older late-type stars \citep{fei05} can be exploited 
to identify young stars and separate them from foreground stellar 
contaminants. Contamination from background AGN can also be accounted for, at least
in part, in a similar fashion, since AGN will have harder X-ray
emission spectra, and be subject to even larger extinction with respect
to members.

In this paper, we focus on the {\em Chandra} CCD-resolution X-ray spectra of low-mass ($\la 2 M_\odot$) Cygnus\,OB2 members to characterize their X-ray emitting plasma, the X-ray extinction due to the intervening interstellar matter, and finally estimate their X-ray fluxes and luminosities. 
Most of our sources are detected with very few counts
and in our analysis we will take advantage of the results of
\citet{alb16}, who use Monte Carlo methods to perform a
statistical assessment of the X-ray spectral fitting procedures in the
low-count regime. We note that the bright spectra of O- and B-type stars, 
as well as the WR stars in Cygnus\,OB2, are discussed in more
detail by \citet{rau15}, while the X-ray properties of intermediate-mass 
stars are presented by Drake et al. (2018b, this issue).

This paper is organized as follows: \S\,\ref{sect:data} describes the
X-ray dataset, summarizes the source classification in terms of
membership, and presents the average spectral features by source class.
The spectral analysis and the methods employed are presented in \S\,\ref{sect:specanal}.
In \S\,\ref{sect:abund} we present how the average abundances of some of the elements in the X-ray emitting plasma were constrained exploiting our large source sample. 
Section \ref{sect:discussion} discusses the implications of our results with regard to the properties of coronal plasma, coronal abundances, and the relation between X-ray and optical extinction toward our Cygnus\,OB members. 
Finally, all the results are briefly summarized in \S\,\ref{sect:summary_conclusions}.

\section{Data and Sample Selection}
\label{sect:data}

Our starting point is the X-ray source catalog and photon extraction
products obtained by \citet{wri14a} using the {\it ACIS Extract} analysis
software \citep{bro10}. In short, using a variety of techniques suitable
for the analysis of multiple, overlapping but non-aligned {\em Chandra} ACIS
observations, source and background photon lists were obtained for each
of the 7924 detected X-ray sources, along with appropriate instrumental
response files. In the great majority of cases, photons for a given
source were collected from several observations, each with a different
effective exposure time, point spread function, background and spectral
response. The {\it ACIS Extract} analysis merges all the {\em suitable}
observations to produce a single average source (and background)
spectrum with associated instrumental response. 

We analyzed these time-averaged spectra for the full sample of 7924
sources. Most of our discussion will, however, be focused on low-mass
Cygnus\,OB2 members. The X-ray spectra of the 100 or so known massive,
O- and B- and WR-, Cygnus\,OB2 stars were also analyzed in the same
way, but the reader is referred to \citet{rau15} for a more in-depth
analysis and discussion of their X-ray properties\footnote{Four stars
were not discussed by Rauw et al.\ because of strong pile-up: CygOB2 \#8a
(cxo\_id 4607), \#5 (2197), \#9 (4377), and \#12 (2926). They are included in our analysis but results are highly uncertain.}. Excluding these high
mass stars, the number of {\em extracted} source photons in the 0.5-8.0\,keV band, 
including the background contribution, range from
3 to 3870. In the following, we will also consider a sub-sample of
{\em bright} X-ray sources, i.e.\ the 2805 sources found with $>$20 net (i.e. background subtracted) counts, to which we
will limit our spectral fitting analysis. 

The background contribution
within the source extraction regions can be significant for fainter
sources, which are the great majority: for 33\% of all sources (46\% of
sources with $<$20 net counts), background photons are more numerous
than source photons. Only for 7.2\%  of the sources (2.0\% of those with $<$20 counts) does the
background contribute less than 10\% of the net source counts. These
large background contributions are partly due to the mapping strategy of
the survey, which resulted in all sources being observed both close to
on-axis, where the PSFs and extraction regions are small, and off-axis, where the PSFs, extraction regions, and hence, background contributions are commensurately larger. The merged source and background photon lists and spectra 
provided by {\it ACIS Extract} are produced by merging the data from the available observations in such a way as to maximize the signal-to-noise ratio \citep[see][for details]{bro10}.


In the following, we will make use of optical and IR data for our X-ray
sources. The cross-identification process and the adopted optical/IR
catalogs are described by \citet{gua15a}. When using optical/IR
data, or derived quantities such as stellar masses, we will only
consider the 5418 X-ray sources with a single optical/IR likely
counterpart. 
These analyses will therefore exclude the 2423 X-ray sources (31\%) with no plausible optical/IR counterpart and the 83 with more than one. However, when only sources with $>$20 counts are considered, just 10\% are thereby excluded.
We adopted the source classification from \citet{kas18}, 
mostly based on a Bayesian analysis of optical, IR, and X-ray data. In
particular, excluding sources with ambiguous optical/IR counterparts,
the fraction of Cyg\,OB2 members, foreground, and background sources
are 78\%, 6\%, and 16\%, respectively (86\%, 6\%, and 8\% for sources
with $>$20 net counts).


\subsection{Spectral features by source class}

\begin{figure}[!h!]
\epsscale{1.2}
\plotone{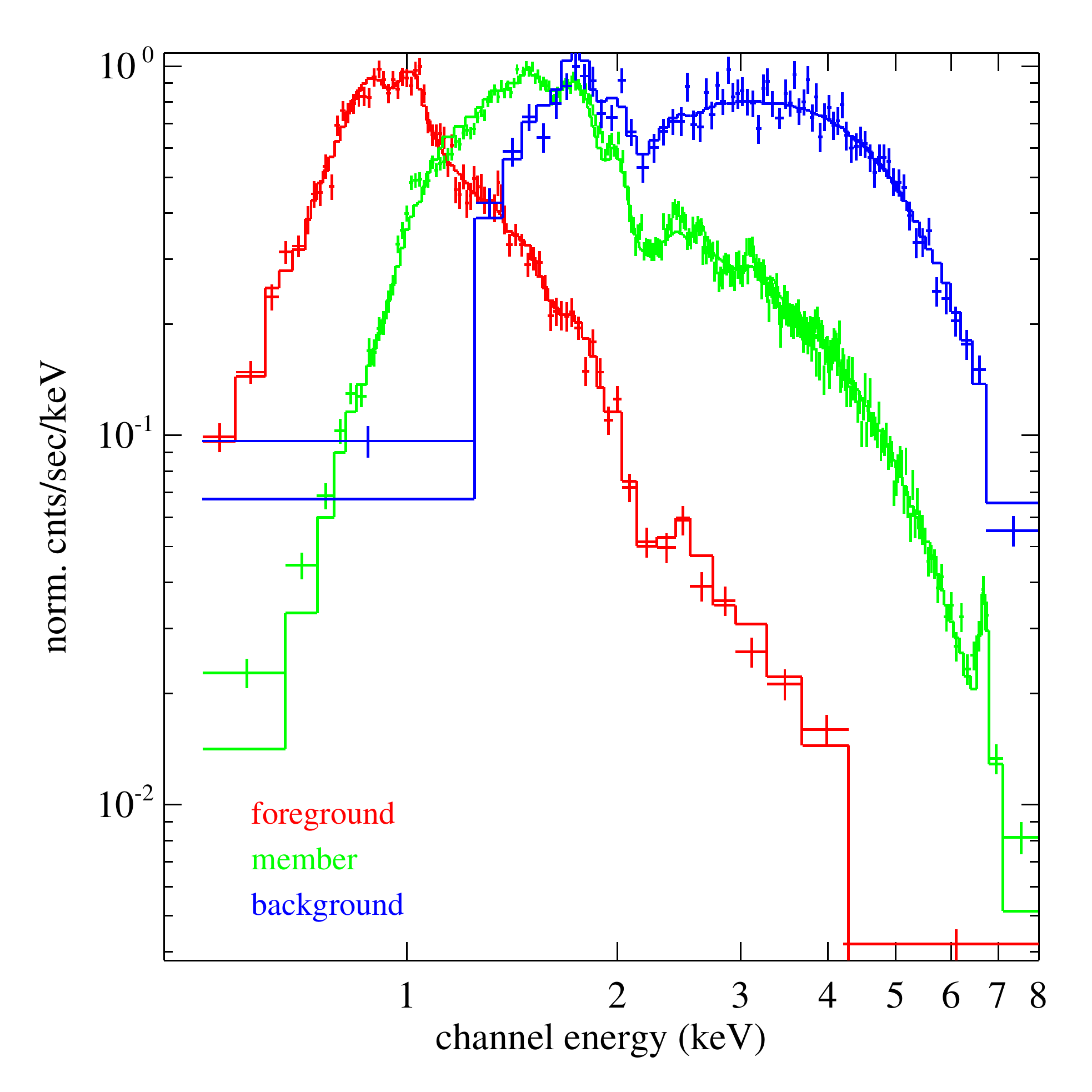}
\caption{Average source spectra for X-ray sources detected with $>$20 net counts, associated with stars of different optical/IR/X-ray classes. The spectra of foreground stars and Cygnus\,OB2 members, in red and green, respectively, are fitted here with absorbed isothermal models ({\sc tbabs$\times$apec}). The average spectrum of background sources, in blue, is instead fitted with an absorbed power-law model ({\sc tbabs$\times$pow}). \label{fig:avsp_class}}
\end{figure}

Figure \ref{fig:avsp_class} shows the average X-ray spectra\footnote{The sum of all spectra divided by the total exposure time, computed with the {\sc combine\_spectra} tool in CIAO \citep{fru06}} for sources classified as low-mass
Cyg\,OB2 members, foreground stars, and background objects with $>20$ net
counts. As expected, foreground stars (mostly unabsorbed main-sequence coronal sources) have X-ray spectra that are softer than those the young low-mass
Cygnus\,OB2 members (absorbed and hotter coronal sources), which are, in
turn, significantly softer than those of background sources (even more absorbed
and intrinsically harder AGNs with non-thermal spectra). These observations support the use of
X-ray spectral characteristics for the assessment of membership, as performed
by \citet{kas18}.
	
\begin{figure}[!t]
\epsscale{1.2}
\plotone{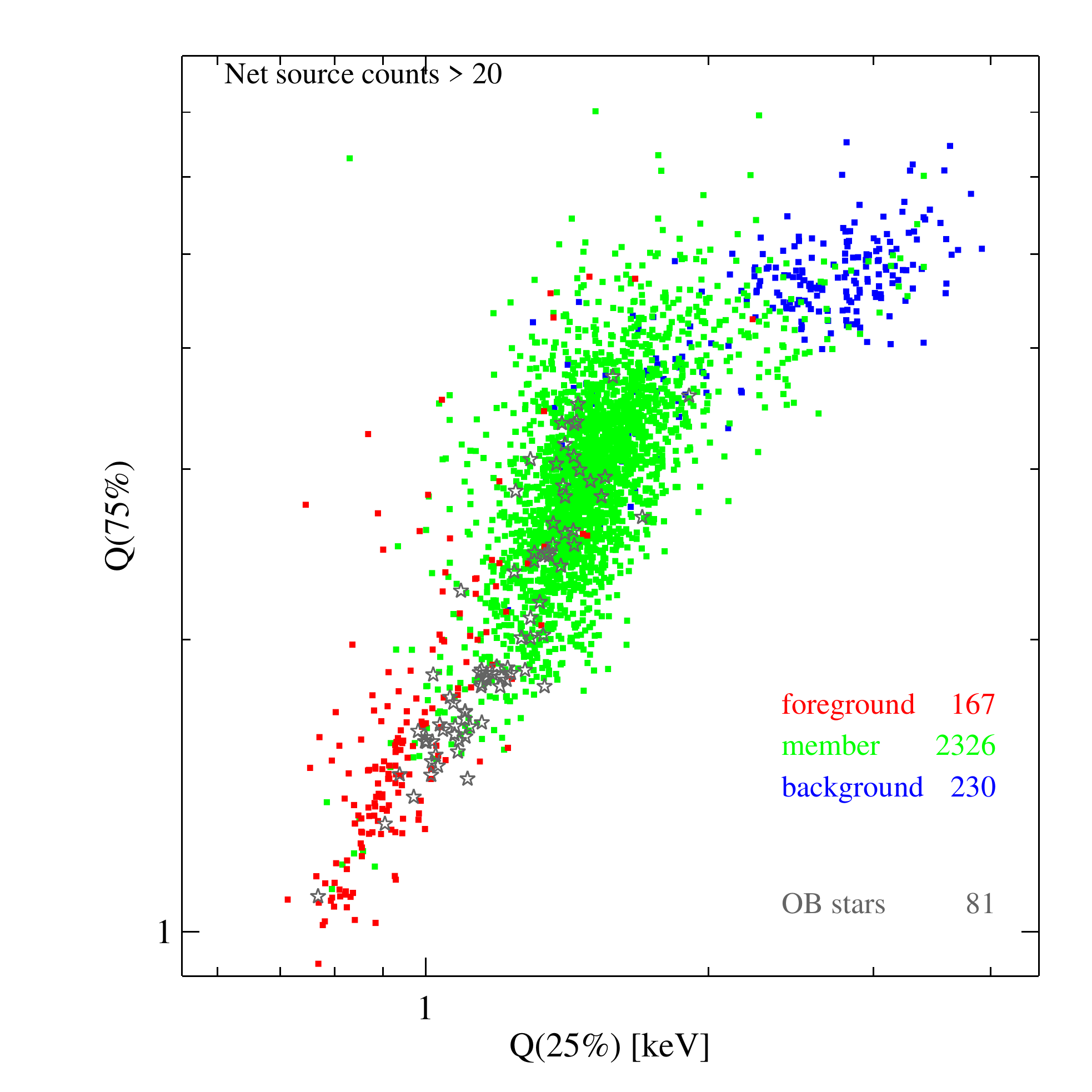}
\caption{$Q_{75\%}$, vs. $Q_{25\%}$ for sources with $>$20 net counts. X-ray sources associated with Cygnus\,OB2 low-mass members, foreground, and background stars are indicated by green, red, and blue squares, respectively. Gray star symbols indicate O- and B-type stars. \label{fig:Q75_Q25}}
\end{figure}

Another, more quantitative, way to show the difference between the three
classes is to look at the quantiles of the energy distribution of
detected photons, $Q_{X\%}$, where $X$ is the fraction of source photons
with energy $<Q_{X\%}$ \citep[e.g.][]{hon04}. The actual calculation of quantiles and their
relation with physical plasma parameters will be discussed in
\S\,\ref{sect:reslt20}. Figure\,\ref{fig:Q75_Q25} shows, for sources with $>20$ net
counts, the 75\% quantile, $Q_{75\%}$, plotted against the 25\% quantile,
$Q_{25\%}$. Green, red, and blue symbols identify the low-mass members, foreground and background stars, respectively, while O- and B-type stars, all presumed to be members of Cyg~OB2, are plotted with gray symbols.
Clearly,  $Q_{75\%}$ and, even more, $Q_{25\%}$ have different
distributions for the three membership classes. As for the OB stars, some fall in the main body of the low-mass members, though a significant fraction have softer spectra, as expected for single O-type stars or systems with only one dominant supersonic wind source \citep{fel97,rau15}.
The harder spectra may instead originate from the coronae of spatially unresolved low-mass companions, which especially for late B type stars may dominate in X-rays,  or from high-mass binaries or multiple systems in which two or more components have supersonic winds that collisionally interact.

\section{Spectral analysis}
\label{sect:specanal}

We here try to constrain the X-ray characteristics of the sources detected in our {\em Chandra} Cygnus\,OB2 survey, such as the temperature of the coronal plasma (for low-mass members of the region), the column density of the intervening absorbing gas, and the intrinsic absorption-corrected X-ray flux.


\subsection{Methods}

The low spectral resolution of ACIS-I pulse height spectra ($\sim$150~eV at 1.5\,keV\footnote{http://cxc.harvard.edu/proposer/POG}) and its complicated
spectral response prevent the direct measurement of most intrinsic
features (e.g.\ line and continuum fluxes) even for high S/N spectra.
Theoretical spectral models of emission and absorption, convolved
through the instrumental response,  must instead be fitted to the
observed spectra in order to characterize the emission and line-of-sight absorption, and
to estimate intrinsic X-ray fluxes. For this purpose we have used the
XSPEC package \citep{arn96} version 12.9 and the models provided therein.
Specifically, we assume that the emission spectra can be described by
the {\sc apec} or {\sc pow} models, the former describing emission from
isothermal optically thin plasma, as expected from stellar coronae, and
the latter a simple power-law spectrum, as expected from extragalactic
AGNs. These models were multiplied by the {\sc tbabs} photoelectric
absorption model to take into account absorption from interstellar
and circumstellar material along the line of sight.

These models, with just one emission component, are the simplest
possible for coronal and extragalactic sources, respectively. However,
as we will see, they are adequate for the overwhelming majority
of our sources. For Cygnus\,OB2 members, this is at odds with results
obtained for young stars in other regions, for which, when enough
signal is available, two isothermal components are often required to provide an adequate 
fit to the emission from what is intrinsically a multi-temperature plasma. This can
likely be ascribed to two factors: $i$) the higher than average
extinction suffered by the stars in Cygnus\,OB2 with respect to other regions (median $A_V$=5.4\,mag \citep{wri15b}, vs. $\lesssim 1$\,mag for, e.g., the ONC or NGC\,2264),
effectively absorbing the lower-energy photons from the cooler thermal
components, up to $\sim$1\,keV (c.f Fig.\,\ref{fig:avsp_class}); $ii$)
the low number of photons detected for most sources.

As noted above, we analyzed the X-ray spectra of all sources with $>$20
counts. The statistical uncertainties associated with such low numbers of counts implies rather large uncertainties on
best-fit model parameters and X-ray fluxes. Care has been taken to
estimate these uncertainties. \citet{alb16} provides a
full discussion of how the uncertainties of best-fit parameters depend on the source photon counts,
background, and spectral parameters. For the fainter Cygnus\,OB2
members, we tried to limit the uncertainties on fluxes by constraining
the spectral parameters $N_H$ and kT to the rather limited ranges
spanned by the brighter members.

For fainter sources, with $<$20 counts, the majority in our X-ray
catalog, we adopted a simpler approach, discussed in \S\,\ref{sect:reslt20}, based
on quantiles of the energies of detected photons.


\subsection{Sources with $>20$ counts}
\label{sect:resgt20}


Prior to spectral fitting, we binned the spectra so that the first
usable energy bin starts at 0.5\,keV, the last ends at 8.0\,keV, and
each bin has a SNR$\gtrsim$1. \citet{alb16} found
that this choice of SNR gives optimal results\footnote{For the binning
we used IDL (Interactive Data Language) and a suitable routine within the ACIS-Extract analysis package ({\sc
group\_bins\_to\_snr}) that computes the SNR per bin using Gehrels'
formula for the uncertainties in source and background counts.
}. Since
the SNR per bin depends on the background in that bin, this procedure
results in bins with a variable number of source counts: depending on
the source the average bin contains  between $\sim$3 and $\sim$120
photons (4.4 on average).  The spectra for 30 sources, 27 of which are associated with members, turned out to have insufficient bins for useful spectral fits and were discarded from the fitting analysis.


Spectral fits of sources with $>$20 net detected counts
were performed with the neutral hydrogen column density, $N_H$, within the {\sc tbabs} model, and the plasma temperature, $kT$, within the 
the {\sc vapec} model, as free parameters. The same spectra were also fit with absorbed power-law models, allowing, again, the neutral hydrogen
column density to vary together with the power-law slope, $\Gamma$, within the {\sc pow} model.
For both emission models, {\sc vapec} or {\sc pow}, the normalizations were kept fixed, since they were substituted, as free fit parameters, with the unabsorbed X-ray fluxes in the 0.5-8.0\,keV band, $F_X$, using the {\sc cflux} convolution model. This has
the advantage of allowing a straightforward estimation of uncertainties
on best-fit unabsorbed flux values. The chemical abundances of the X-ray emitting plasma were adopted from \citet{mag07}, derived from YSOs in the Orion Nebula
Cluster, and were kept fixed in this stage of the fitting process\footnote{The abundances for the absorption model were instead kept at the solar values listed by \citet{and89}}. 
This assumption will then be tested in \S\,\ref{sect:abund} where we will derive  {\em average} abundances largely consistent with \citet{mag07}. Furthermore, in view of the possible dependence of abundances on activity level and/or spectral type (c.f. \S\,\ref{sect:disc_abund}), we checked that this choice, over that of a different and reasonable abundance set, has negligible effects on the estimated X-ray source flux, which is arguably the most important product of our spectral fitting, and small effects on $N_H$ and $kT$ (on average $<10^{21}$\,cm$^{-2}$ and $<$0.1\,keV, respectively). We arrive at these conclusions in Appendix\,\ref{sect:abund_biases} comparing the isothermal fits with \citet{mag07} abundances, described below, with similar fits using two very different and ``extreme'' abundance sets, i.e. those derived for V410\,Tau and SU\,Aur by \citet{tel07a} with high-resolution X-ray spectra.


The Cash statistic (C-stat; \citealt{cas79}) was adopted as the fit statistic, while the Pearson
$\chi^2$ was employed as a test statistic, i.e. to estimate,  through the
{\sc goodness} XSPEC command, how well the best-fit spectral model fits
the data.  Uncertainties (1$\sigma$) on our three parameters, $N_H$,
$kT$ (or $\Gamma$), and $F_X$ were obtained using the {\sc error} XSPEC
command. In a very small number of cases, the calculation of
uncertainties failed, although
this is fortunately rarer for the flux than for the other two
parameters. For the final choices of spectral models, upper and/or lower
flux uncertainties are missing for only eleven sources, none of which was  classified as a Cygnus OB2 member.


Adopting absorbed isothermal models with unconstrained $N_H$ and $kT$ resulted in intrinsic flux estimates of low count sources being very
uncertain. For example, among the 531 low-mass members with 20-25 net
counts,
54\%(14\%) have 1$\sigma$ flux uncertainties larger than 0.3(0.8)\,dex. This is mainly the
result of the large (and correlated) uncertainties on $N_H$ and $kT$. In
order to establish reasonable flux estimates for ensuing studies, we
have tried to reduce uncertainties by making reasonable assumptions on
the range of true values for cluster members, based on the
better-defined estimates obtained for the brightest sources.

\begin{figure*}[!th!]
\epsscale{1.17}
\plottwo{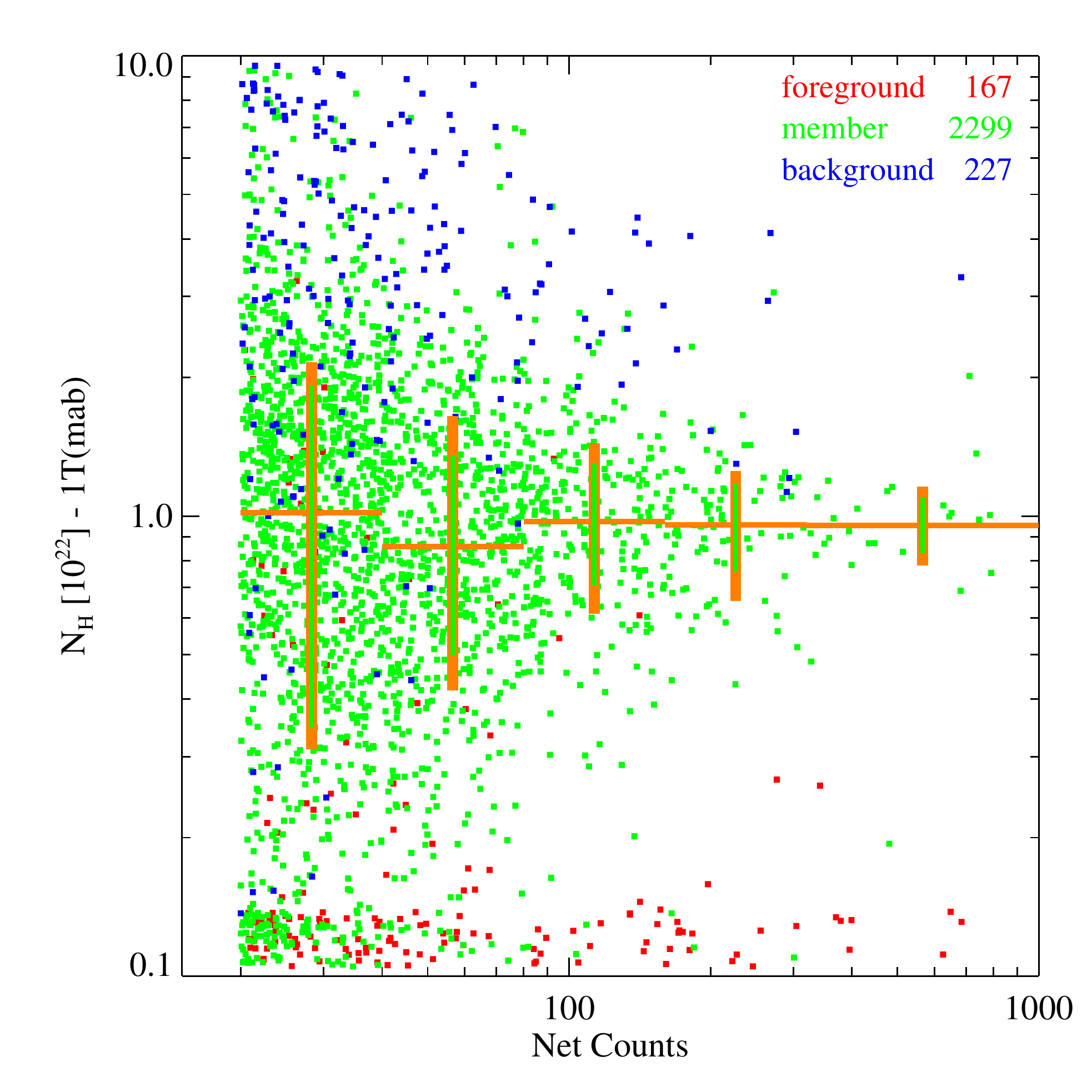}{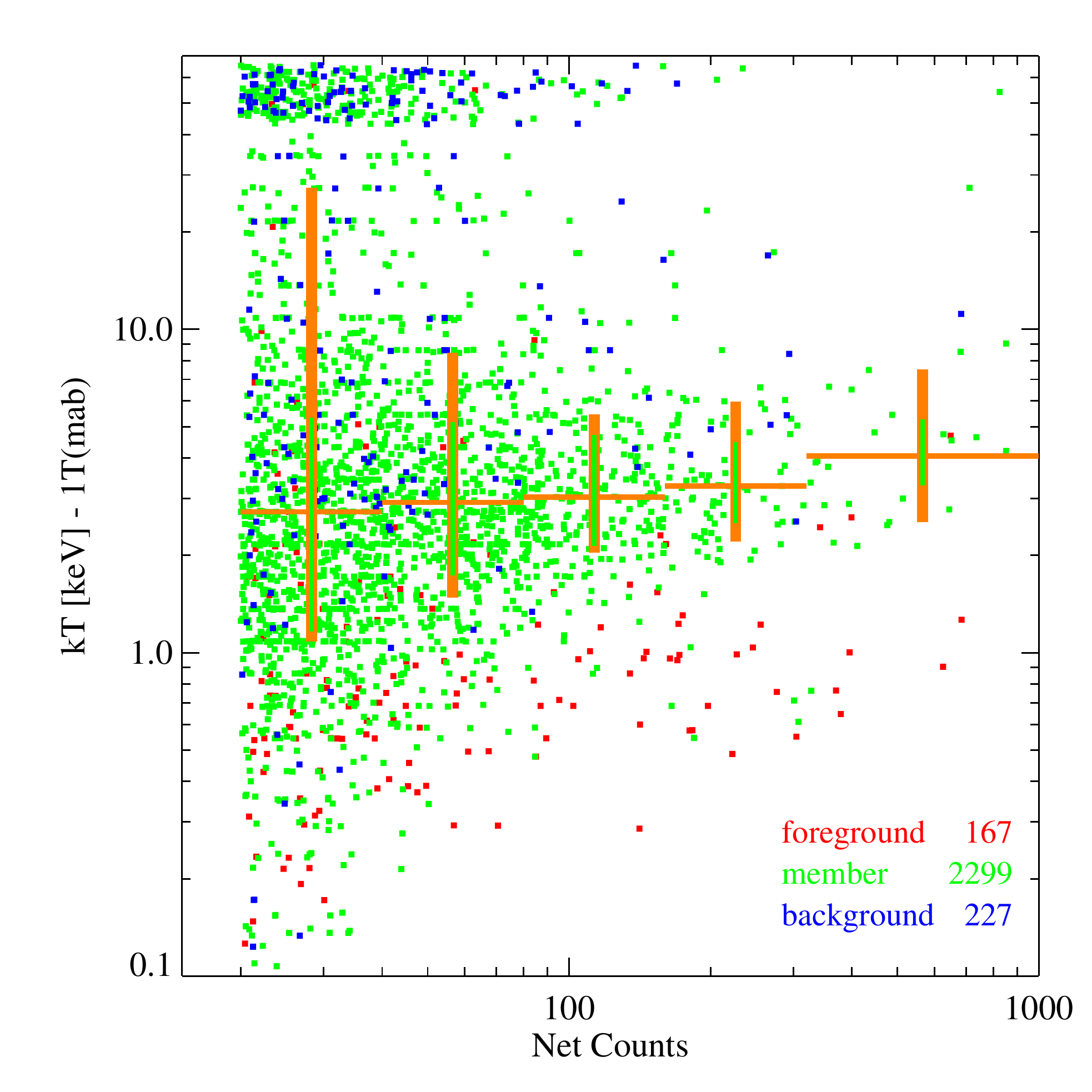}
\caption{$a)$ Best fit $N_H$ from isothermal fits, vs. source counts.
X-ray sources are color-coded according to their optical/IR class.
Values beyond the plotting limits are plotted close to the respective
axes, resulting in an artificial accumulation of points.  The large
orange crosses indicate the median and $\pm 1\sigma$ quantiles of the
distribution of $N_H$ for members in selected ranges of net counts,
marked by the horizontal bars. The lengths of the green bars within the
thick orange ones  represent the median of the upper and lower
statistical uncertainties on $N_H$ for the same samples.   $b)$ The same
as $a)$ for the best fit $kT$. \label{fig:NH_kT_counts}}
\end{figure*}

Figure \ref{fig:NH_kT_counts} shows the trends with source statistics
of $N_H$ and $kT$ from unconstrained fits. Sources are color coded
according to their member/foreground/background classification in
\citet{kas18}. Also shown are median error bars (1$\sigma$,
green segments), median parameter values and 1$\sigma$ scatter (thicker
crosses in orange), all computed, exclusively for members, in five
different source count intervals. For both quantities, and in particular for $N_H$, the uncertainties
seem to explain much of the scatter among members, indicating that $i)$
the real ranges of $N_H$ and $kT$ spanned by members are probably rather
small in comparison to the measurement error, and, $ii)$ that there is little evidence for a dependence of
these ranges on source counts (or intensity). The great majority of
low-mass members with high counts have $N_H$ between 0.6 and
$1.5 \times 10^{22}$cm$^{-2}$, with a median value, for all low-mass memers, of $0.95 \times 10^{22}$cm$^{-2}$.
Likewise, the median $kT$ of low-mass members is $\sim$2.9\,keV.
In this case, we may notice a shallow trend in the median $kT$ with counts,
consistent with previous reports indicating a direct relation between
mean activity levels and plasma temperature  \citep[e.g.][]{pre05a,tel07}. A high-$kT$ tail is also apparent in the low-counts bins, likely due to the hot plasma produced during magnetic flaring, and possibly more frequently at the faint luminosity end because of a detection bias. Following these considerations, it seems reasonable to assume that the characteristic temperature\footnote{I.e., the temperature that best represents the absorbed spectrum as fit by an isothermal model.} of the vast majority of Cygnus OB2 members ranges between 2.0 and 10.0\,keV.

In order to try to limit the
uncertainties on the intrinsic X-ray fluxes of Cygnus\,OB2 members, while at
the same time trying not to impose too strong a bias, we chose a 
preferred spectral model using the following procedure. We start by adopting
the unconstrained 1T spectral fits. For members only, if the 1$\sigma$
{\em range} of allowed fluxes is larger than 0.6\,dex, i.e. if the
one-tailed uncertainty is greater than a factor of $\sim$2,  we then
considered a 1T fit in which $N_H$ is constrained between 0.6 and
$1.5\times10^{22}$cm$^{-2}$ (we will refer to this model as $\rm 1T_{NH}$). We adopt this latter fit only if
statistically acceptable ($P_{null}>1\%$ or larger than that of
the current choice, i.e. the unconstrained 1T fit) and the associated
1$\sigma$ {\em range} of fluxes is $<$0.6\,dex (or $<$1/10th that of the
unconstrained 1T fit). Finally, if the 1$\sigma$ {\em range} of allowed
%
%
%
%
fluxes is still larger than 0.6\,dex, we considered the 1T fits in
which, in addition to $N_H$, constrained as above, we also constrained
$kT$ to lie between 2.0 and 10.0\,keV ($\rm 1T_{NH,kT}$). This latter fit was
preferred to the current choice (1T or $\rm 1T_{NH}$) only if statistically
acceptable ($P_{null}>1\%$ or anyway larger than that of the current
choice) and the associated 1$\sigma$ {\em range} of fluxes was
$<$0.6\,dex wide (or $<$1/10th that of the current choice). 
As a last step, we examined by eye the X-ray spectra of sources
associated with Cygnus\,OB2 members and having the poorest spectral
fits, as well as those for which the automatically assigned parameters
were physically unreasonable (e.g.\ excessively low kT or very high fluxes).
In 12 cases, new spectral models were assigned. In 11 of these cases, a
second isothermal component was added to the emission model to improve the quality of the fit.

This procedure results in the adoption of 1T/$\rm 1T_{NH}$/$\rm 1T_{NH,kT}$ fits for
70.8/18.8/9.3\% of the 2362 low-mass members with $>$20
counts\footnote{1.2\% of these sources have no fits, because the binned
spectra have too few degrees of freedom}. These fractions become
94.5/4.4/1.1\%, for the 840 low-mass members with $>50$ counts and 
99.0/0.7/0.4\% for the 287 low-mass members with $>$100 counts. In spite of this
procedure, 2\% of the low-mass members with $>$20 counts (and mostly $<$50
counts) remain with a 1$\sigma$ flux range $>$0.6\,dex.

\begin{figure}[t]
\epsscale{1.2}
\plotone{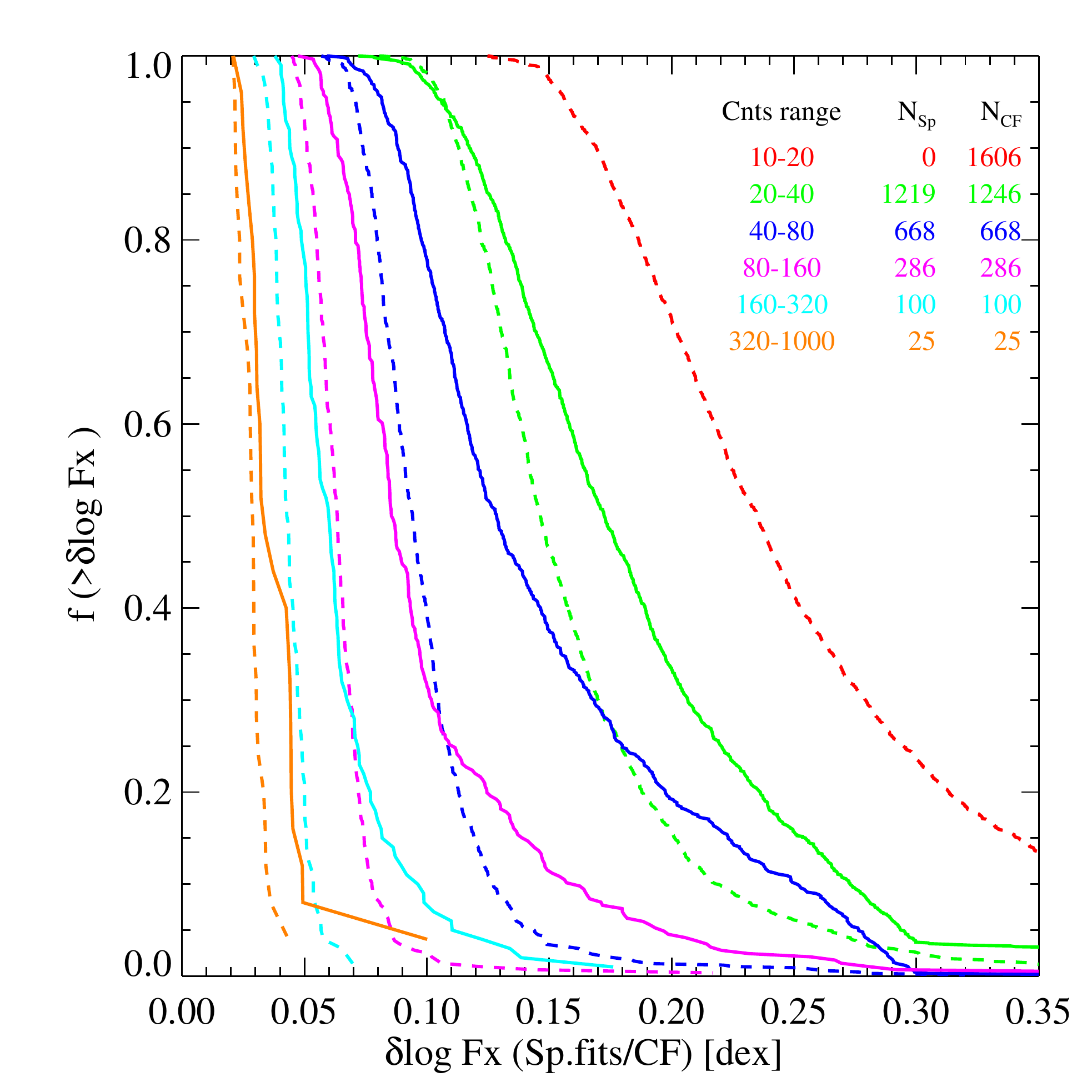}
\caption{Cumulative distribution of uncertainties on the unabsorbed X-ray flux for {\em members} in different ranges of detected counts. Solid curves refer to uncertainties derived from spectral fits (\S\,\ref{sect:resgt20}). Dashed curves refer to the uncertainties on the fluxes derived using observed to intrinsic flux conversion factors (\S\,\ref{sect:reslt20}). 
In all cases the uncertainties are the mean of the upper and lower uncertainties. Count ranges and respective numbers of sources in the two samples are listed in the legend. \label{fig:DFx_counts}}
\end{figure}

Figure\,\ref{fig:DFx_counts} shows the distributions of the final flux uncertainties (mean
between upper and lower uncertainties) as a function of detected counts. The full results of our
%
%
%
%
spectral fitting are reported in Table~\ref{tab:spectral_fits}. Columns
1-3 list, for each X-ray source, the source identifier, the net counts
detected in the 0.5-8.0\,keV band and the membership class from Kasyhap
et al. (2018). Columns 4-7 indicate, for thermal model fits, the
adopted emission model (i.e. whether one or two thermal components were
needed, and whether $N_H$ and $kT$ were constrained), the null probability
of the fit, and the values and uncertainties of the best fit parameters:
$N_H$, $kT$ (the emission measure-weighted mean values for 2T models), and unabsorbed flux. As
mentioned, we have also fitted all of our spectra with absorbed
power-laws, suitable for background AGN. Unfortunately, our ACIS spectra
are not adequate to distinguish between the intrinsic emission spectrum
of a star and that of an AGN for the majority of sources, both because of their low signal-to-noise ratio, and because Cygnus\,OB2 members have relatively hot coronae producing
spectra that are more similar to power-laws than those of cooler
coronae. Results of the power-law fits are also reported in Table~\ref{tab:spectral_fits}
(col. 8-11). Table\,\ref{tab:2t_fits} provides the spectral
parameters for the few X-ray sources needing a second thermal component.

\begin{figure}[!t!]
\epsscale{1.2}
\plotone{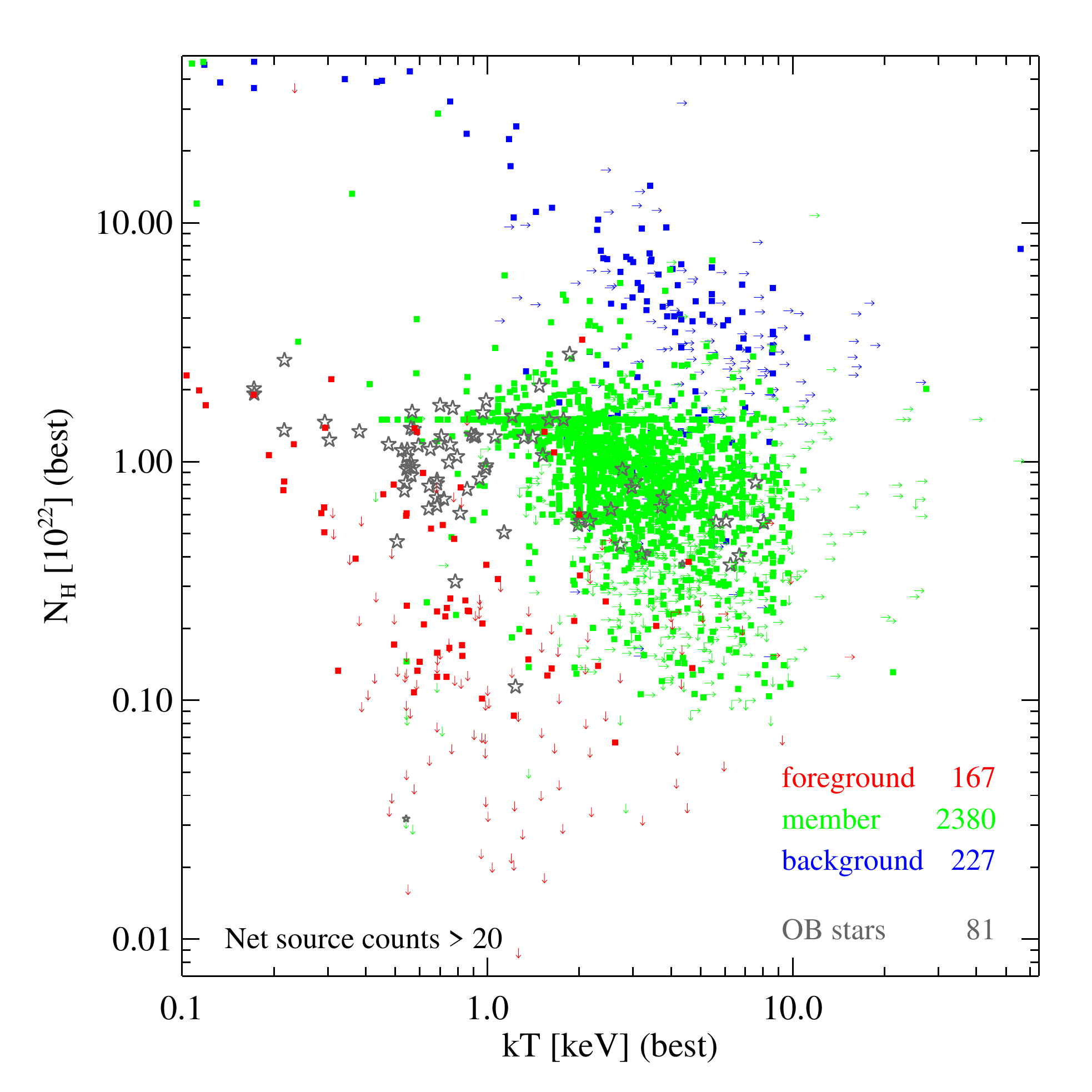}
\caption{Final $N_H$ vs. $kT$ scatter plot. Symbols are color-coded according to the optical/IR classification of sources, as noted in the legend. $N_H$ values below the plotting limits and compatible with zero are plotted as upper limits at their upper 1$\sigma$ uncertainty limit. Similarly, $kT$s with unconstrained upper bounds are plotted as lower limits at their lower 1$\sigma$ limit. \label{fig:NH_kT}}
\end{figure}

Figure\,\ref{fig:NH_kT} shows $N_H$ vs.\ $kT$ (average $kT$ for
2-temperature models), color coded according to our membership 
classification. The lowest $N_H$ values and highest $kT$s, for which
only upper and lower limits, respectively, can actually be constrained, are plotted as
arrows corresponding to their 1$\sigma$ limits.

\subsection{Sources with $<$20 counts}
\label{sect:reslt20}

For sources with fewer than 20 detected counts, for which model fitting is impractical, we decided to adopt a different approach based on the spectral information summarized by chosen quantiles of the distribution of detected photon energies.

First we computed, from the ACIS spectrum of each source, 11 quantiles,
$Q_{X\%}$, where $X$, the fraction of source photons with energy
$<Q_{X\%}$, ranges from 0.1 to 0.9 in steps of 0.1 and additionally 
included the values 0.25 and 0.75.
The $Q_{X\%}$ values, and their associated $1\sigma$ uncertainties,
$\delta Q_{X\%}$, were estimated accounting for the background
contribution to extracted source photons. To this end we followed
\citet{hon04} and used their IDL routine (version 1.7).

For similar intrinsic spectra, the quantiles are sensitive to the
absorbing column density. This latter is, in turn, the main factor
determining the reduction in flux from the intrinsic one at the source
to the one observed at the detector (barring flux dilution due to
distance, of course). We thus sought a relation between energy quantiles
and the conversion factor (CF) needed to transform detected photon
fluxes (in ph\,cm$^{-2}$\,s$^{-1}$) into absorption-corrected energy
fluxes (in erg\,cm$^{-2}$\,s$^{-1}$).

\begin{figure*}[t]
\epsscale{1.17}
\plottwo{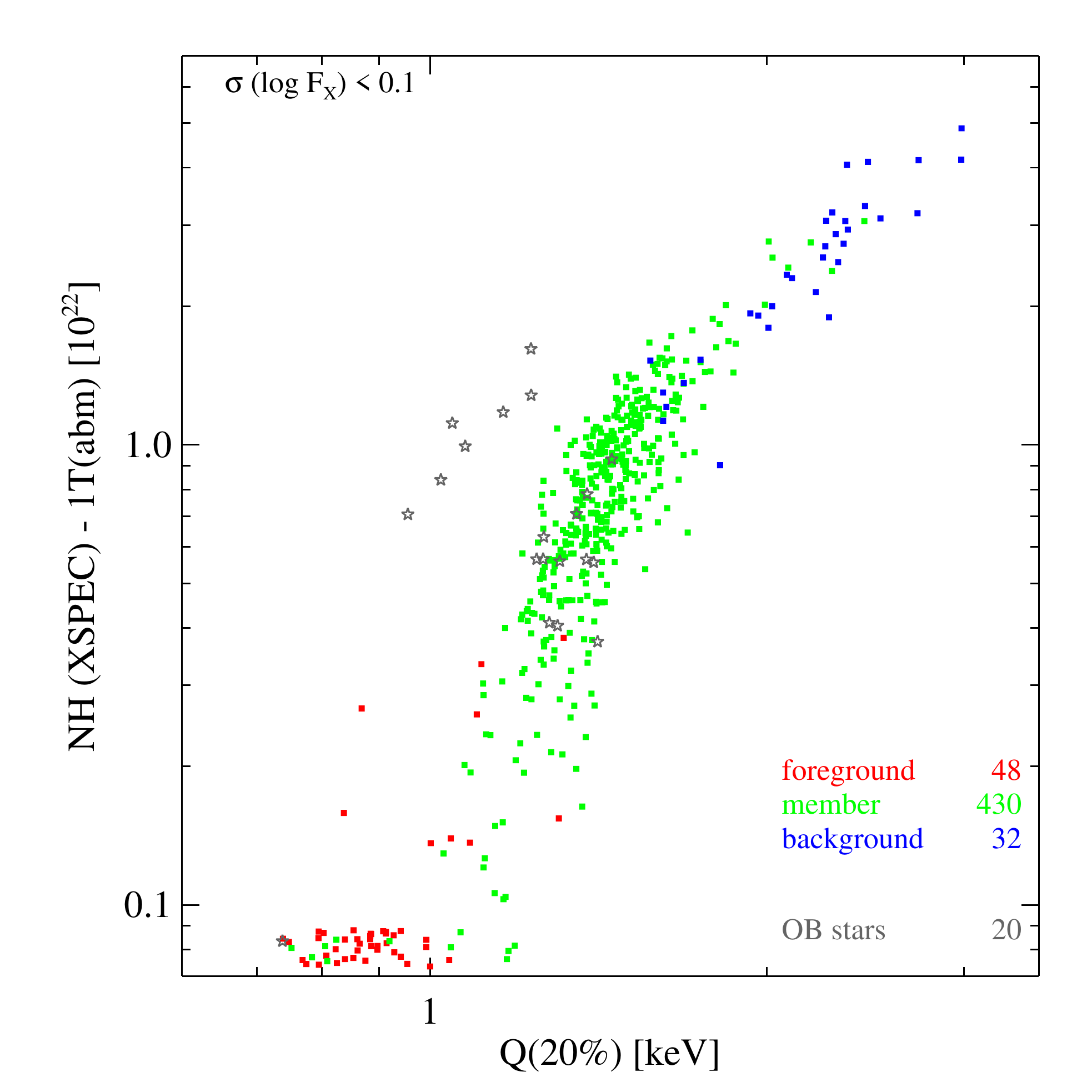}{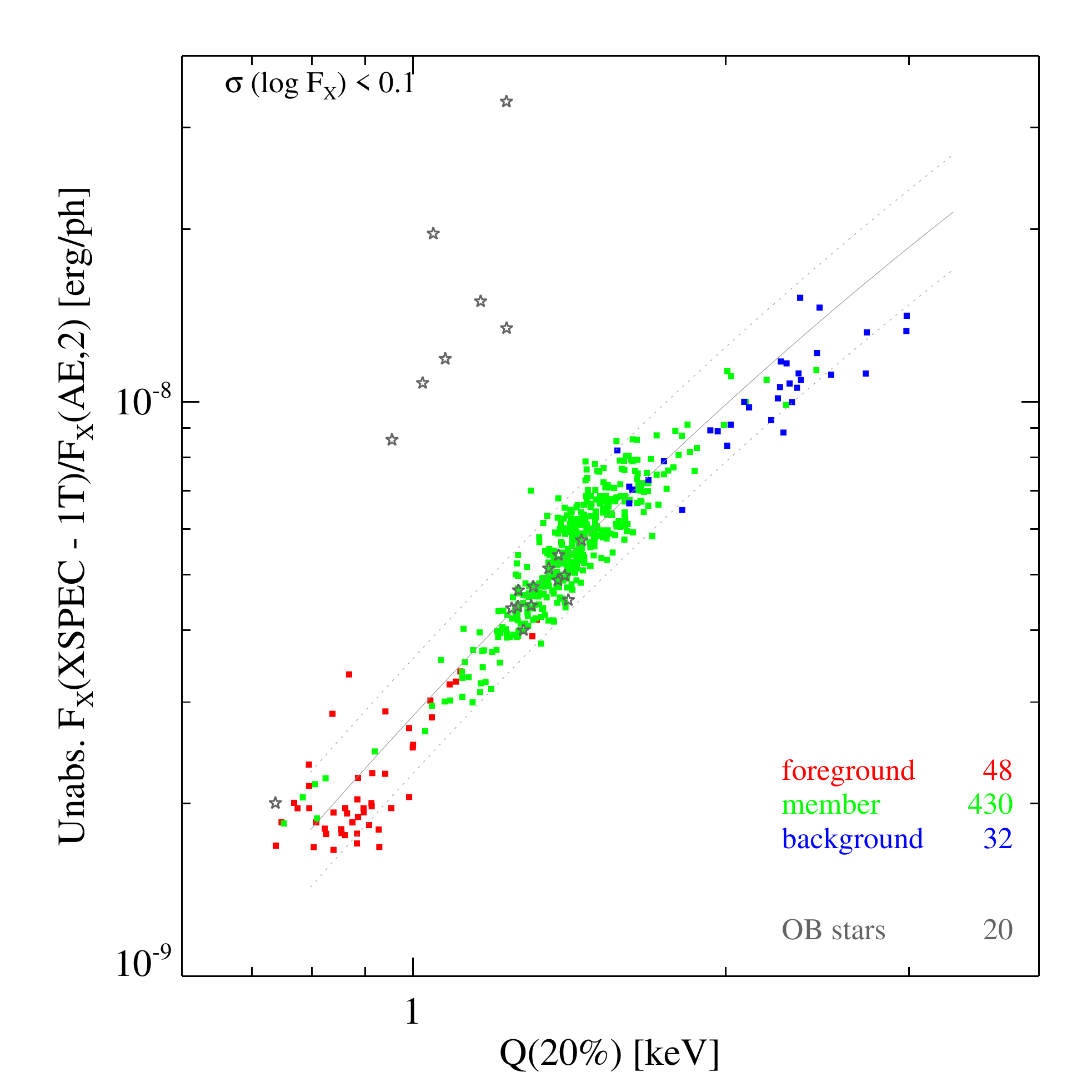}
\caption{[left]: $N_H$, as estimated through spectral fitting vs.\ $Q_{20\%}$, the 20\% quantile of the photon energy distribution. Symbols are color-coded according to the optical/IR classification, as indicated in the legend. Only stars with mean flux uncertainties $<$0.1\,dex are plotted. $[right]$: Ratio of unabsorbed energy flux to observed photon flux plotted as a function of $Q_{20\%}$. Sample and symbols as in the left-hand panel. The solid line indicates a quadratic fit to the all points except O/B-type stars, and the dashed lines are the same best-fit shifted by $\pm$0.1\,dex.   \label{fig:NHCF_Q20}}
\end{figure*}

We first investigated the sensitivity of each $Q_{X\%}$ to both of the
X-ray spectral parameters, $N_H$ and $kT$. In Fig.\,\ref{fig:NHCF_Q20}a we
plot, for example, $N_H$ vs.\ $Q_{20\%}$ for sources for which the
$1\sigma$ uncertainty on the unabsorbed flux from spectral fits is
$<0.1$\,dex. Points are color-coded according to optical/IR class. We
find generally reasonable results: $N_H$ correlates better with {\em
low} quantiles, e.g.\  $Q_{10\%}$ or $Q_{20\%}$, while the quality of the
correlation decreases as we move to {\em higher} quantiles. This is
readily understood as low quantiles are good proxies of the low-energy
cutoff due to absorption. $kT$, on the other hand, correlates better
with higher percentiles, although all quantiles saturate at high $kT$
values. As for the {\em conversion factor}, $CF$, i.e.\ the ratio between
detected photon flux and the absorption-corrected energy flux, as
estimated through spectral fitting, Fig.\,\ref{fig:NHCF_Q20}b shows the
correlation with $Q_{20\%}$ (in the same sample as in Fig.\,\ref{fig:NHCF_Q20}a,
i.e.\ sources with $\sigma(\log F_X)<0.1$\,dex). Disregarding O- and
B-type stars (the gray star-shaped symbols), a tight correlation is
observed. In particular, for low-mass Cygnus\,OB2 members, the $1\sigma$
y-axis dispersion of members around the quadratic best-fit
correlation illustrated (solid gray line) is 0.044\,dex, which is 
smaller than the median y-axis uncertainty of 0.075\,dex for the points
shown. Such dispersion about the mean relation is smaller than
that obtained for the same stars using the median photon energy,
$Q_{50\%}$: 0.063\,dex.\footnote{Larger dispersions, but consistent
trends with percentiles are found when loosening the selection criterion on
reliable flux values. Repeating the same correlation studies using all
members with uncertainty on the unabsorbed flux from spectral fits
$<0.5$\,dex (instead of 0.1), we observe that the y-axis dispersions with
respect to the mean relation for $Q_{20\%}$ and $Q_{50\%}$
are 0.142 and 0.163\,dex, respectively.} We conclude that low quantiles are
best suited to defining a relation with conversion factor. This was not
obvious {\it a priori} since, while we have seen that the relation with $N_H$
is indeed better described by the low quantiles, the statistical
uncertainties in their estimates are larger than those on quantile
fractions closer to 50\%.

%

We estimate unabsorbed fluxes from $Q_{20\%}$, the conversion factors derived from the quadratic fit in Fig.\,\ref{fig:NHCF_Q20}b (limiting low values of 
$Q_{20\%}$ to 0.8, corresponding to $\sim$zero absorption), and the photon fluxes provided by {\it ACIS Extract}. Uncertainties on these fluxes were estimated propagating
the uncertainties on $Q_{20\%}$ and those on the observed
photon flux, taken to be of the same relative magnitude as the Poisson
uncertainties on the observed counts.

\begin{figure}[t]
\epsscale{1.2}
\plotone{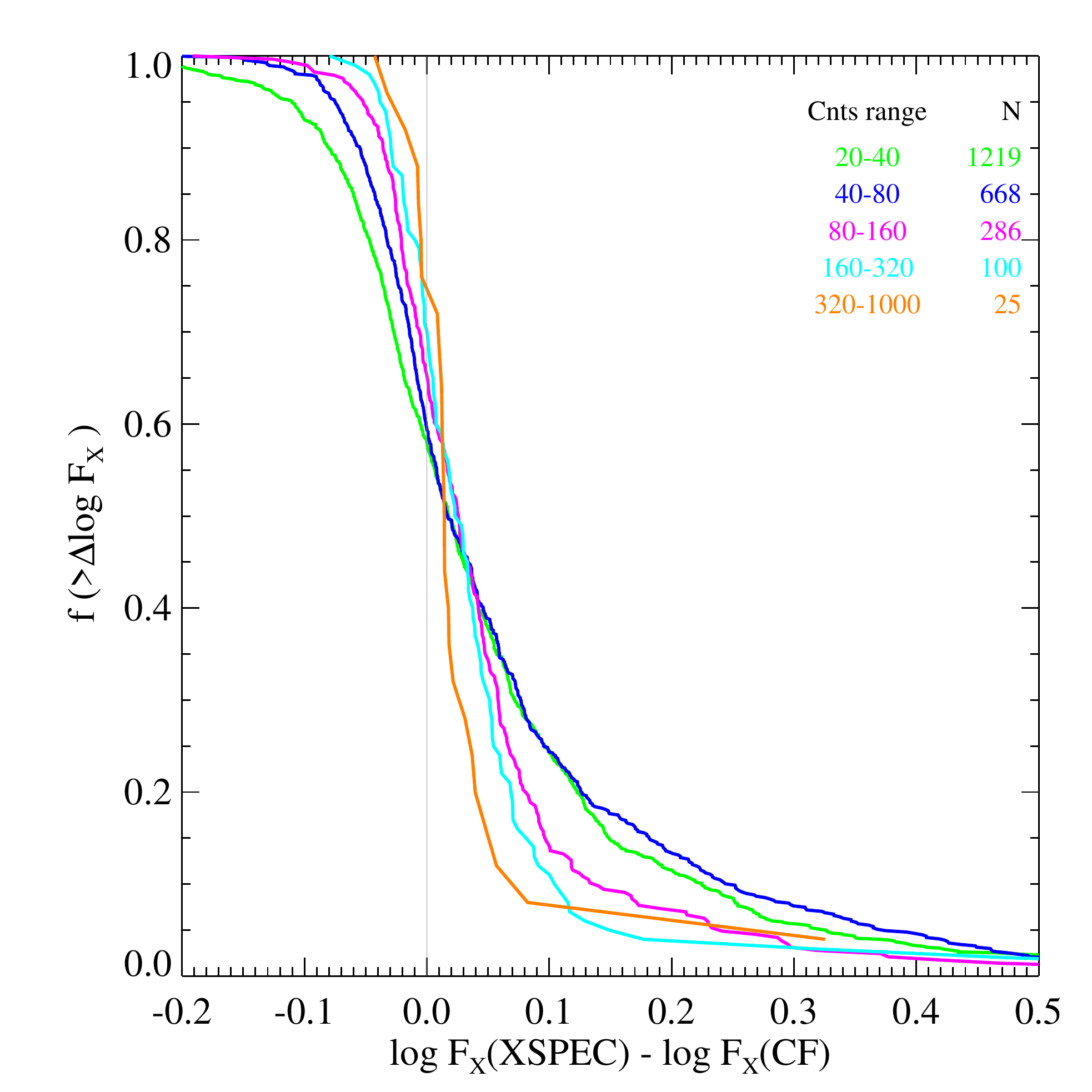}
\caption{Difference between the logarithmic fluxes obtained from the $Q_{20\%}$-dependent conversion factor and those obtained from spectral fits, for low-mass members in different ranges of net counts. \label{fig:FxCF_FxX}}
\end{figure}

Figure\,\ref{fig:FxCF_FxX} shows, in different ranges of net source
counts, the distributions of the differences between the fluxes obtained
from the conversion factors with those obtained in
\S\,\ref{sect:resgt20} from spectral fittings. The median difference
between the two estimates shows a small systematic offset, of order
0.01-0.02\,dex, that is rather constant with source counting statistics. Differences
can, however, become particularly severe and asymmetric at low counts,
likely because of the significant uncertainties on the absorption
correction obtained from spectral fitting, which sometimes leads  to large and
uncertain absorptions. Taking uncertainties into account, we indeed see
that the two estimates are compatible within 1 and 2$\sigma$ for 90\%
and 98\% of the low-mass members, respectively.

Distribution of the flux uncertainties as derived using CFs are shown in Fig.\,\ref{fig:DFx_counts} separately for stars in several count ranges (dashed curves), and compared with those from spectral fits (solid curves). 

\begin{table*}[h]
\begin{center}
\scriptsize
\caption{Results of spectral fits and quantile analysis\label{tab:spectral_fits}}
{
\setlength{\extrarowheight}{0.5em}
\begin{tabular}{lrrrrrrrrrrrrr}
\tableline\tableline
\multicolumn{3}{c|}{Source}&\multicolumn{5}{c|}{Isothermal fit} &\multicolumn{4}{c|}{Power law fit} & \multicolumn{2}{c}{Conversion Factor}\\
Id.&    Counts&          Class&               Model&  n.p&                            nH&                            kT&                      log Flux&  n.p&                            nH&                      $\Gamma$&                      log Flux&                    $Q_{20\%}$&                      log Flux\\
\tableline
                   1&      13.8&           memb&                    &     &                      $_{}^{}$&                      $_{}^{}$&                      $_{}^{}$&     &                      $_{}^{}$&                      $_{}^{}$&                      $_{}^{}$&           $1.61$$ \pm $$0.29$&        $-13.77_{0.21}^{0.21}$\\
                   2&      24.9&           memb&                  1T& 59.9&          $3.18_{1.35}^{2.73}$&              $68.4_{63.5}^{}$&        $-13.29_{0.16}^{0.20}$& 94.4&              $0.97_{}^{4.52}$&            $-0.2_{0.9}^{1.5}$&        $-13.37_{0.14}^{0.13}$&           $2.32$$ \pm $$0.58$&        $-13.21_{0.23}^{0.19}$\\
                   3&      15.3&           memb&                    &     &                      $_{}^{}$&                      $_{}^{}$&                      $_{}^{}$&     &                      $_{}^{}$&                      $_{}^{}$&                      $_{}^{}$&           $2.12$$ \pm $$0.38$&        $-13.01_{0.19}^{0.19}$\\
                   4&      23.2&             bk&                  1T& 43.7&          $5.64_{2.14}^{4.30}$&              $68.4_{64.0}^{}$&        $-13.17_{0.15}^{0.26}$& 45.1&              $2.84_{}^{8.34}$&            $-0.1_{1.2}^{1.7}$&        $-13.33_{0.13}^{0.26}$&           $2.99$$ \pm $$0.69$&        $-13.08_{0.20}^{0.17}$\\
                   5&      18.2&           memb&                    &     &                      $_{}^{}$&                      $_{}^{}$&                      $_{}^{}$&     &                      $_{}^{}$&                      $_{}^{}$&                      $_{}^{}$&           $1.51$$ \pm $$0.36$&        $-13.63_{0.24}^{0.21}$\\
                   6&      22.5&             bk&                  1T& 97.6&          $2.97_{1.16}^{2.69}$&              $68.4_{64.7}^{}$&        $-13.35_{0.14}^{0.24}$& 97.8&          $3.19_{3.16}^{5.86}$&             $0.8_{1.3}^{1.6}$&        $-13.42_{0.15}^{0.48}$&           $2.39$$ \pm $$0.34$&        $-13.27_{0.15}^{0.15}$\\
                   7&       4.4&           memb&                    &     &                      $_{}^{}$&                      $_{}^{}$&                      $_{}^{}$&     &                      $_{}^{}$&                      $_{}^{}$&                      $_{}^{}$&           $0.56$$ \pm $$0.19$&        $-14.85_{0.26}^{0.38}$\\
                   8&       4.7&           memb&                    &     &                      $_{}^{}$&                      $_{}^{}$&                      $_{}^{}$&     &                      $_{}^{}$&                      $_{}^{}$&                      $_{}^{}$&           $1.07$$ \pm $$0.09$&        $-14.58_{0.25}^{0.35}$\\
                   9&       9.6&             fg&                    &     &                      $_{}^{}$&                      $_{}^{}$&                      $_{}^{}$&     &                      $_{}^{}$&                      $_{}^{}$&                      $_{}^{}$&           $0.78$$ \pm $$0.12$&        $-14.41_{0.16}^{0.24}$\\
                  10&      38.8&             bk&                  1T& 98.6&          $4.47_{1.84}^{3.28}$&             $2.8_{1.3}^{9.5}$&        $-12.87_{0.24}^{0.44}$& 96.9&          $6.87_{3.43}^{6.68}$&             $2.5_{1.2}^{1.7}$&        $-12.72_{0.42}^{1.13}$&           $2.45$$ \pm $$0.31$&        $-12.97_{0.12}^{0.12}$\\
                  11&       3.4&           memb&                    &     &                      $_{}^{}$&                      $_{}^{}$&                      $_{}^{}$&     &                      $_{}^{}$&                      $_{}^{}$&                      $_{}^{}$&           $0.71$$ \pm $$0.06$&        $-14.98_{0.28}^{0.43}$\\
                  12&       8.3&             bk&                    &     &                      $_{}^{}$&                      $_{}^{}$&                      $_{}^{}$&     &                      $_{}^{}$&                      $_{}^{}$&                      $_{}^{}$&           $1.52$$ \pm $$0.29$&        $-13.99_{0.25}^{0.27}$\\
                  13&       7.5&             fg&                    &     &                      $_{}^{}$&                      $_{}^{}$&                      $_{}^{}$&     &                      $_{}^{}$&                      $_{}^{}$&                      $_{}^{}$&           $0.95$$ \pm $$0.07$&        $-14.45_{0.19}^{0.25}$\\
                  14&      12.3&           memb&                    &     &                      $_{}^{}$&                      $_{}^{}$&                      $_{}^{}$&     &                      $_{}^{}$&                      $_{}^{}$&                      $_{}^{}$&           $2.06$$ \pm $$0.39$&        $-13.63_{0.21}^{0.21}$\\
                  15&     126.6&             fg&                  1T&  0.4&          $0.21_{0.12}^{0.13}$&             $1.0_{0.1}^{0.1}$&        $-13.21_{0.12}^{0.13}$&  0.4&          $0.45_{0.17}^{0.19}$&             $4.0_{0.5}^{0.6}$&        $-12.85_{0.19}^{0.24}$&           $0.93$$ \pm $$0.02$&        $-13.27_{0.04}^{0.05}$\\
                  16&      79.8&           memb&                  1T& 70.3&          $6.84_{1.79}^{2.66}$&               $13.7_{9.8}^{}$&        $-12.60_{0.12}^{0.21}$& 72.9&         $12.23_{4.30}^{5.93}$&             $1.9_{0.8}^{1.0}$&        $-12.47_{0.26}^{0.53}$&           $3.04$$ \pm $$0.13$&        $-12.59_{0.06}^{0.06}$\\
                  17&      72.6&           memb&                  1T& 76.6&          $1.24_{0.35}^{0.49}$&              $22.7_{16.8}^{}$&        $-13.02_{0.07}^{0.08}$& 77.5&          $1.92_{0.74}^{0.88}$&             $1.5_{0.5}^{0.5}$&        $-13.01_{0.09}^{0.13}$&           $1.66$$ \pm $$0.13$&        $-13.02_{0.08}^{0.08}$\\
                  18&       4.5&           memb&                    &     &                      $_{}^{}$&                      $_{}^{}$&                      $_{}^{}$&     &                      $_{}^{}$&                      $_{}^{}$&                      $_{}^{}$&           $1.40$$ \pm $$0.18$&        $-14.38_{0.25}^{0.36}$\\
                  19&       9.5&           memb&                    &     &                      $_{}^{}$&                      $_{}^{}$&                      $_{}^{}$&     &                      $_{}^{}$&                      $_{}^{}$&                      $_{}^{}$&           $3.52$$ \pm $$0.63$&        $-13.36_{0.19}^{0.22}$\\
                  20&       4.8&             bk&                    &     &                      $_{}^{}$&                      $_{}^{}$&                      $_{}^{}$&     &                      $_{}^{}$&                      $_{}^{}$&                      $_{}^{}$&           $2.58$$ \pm $$0.66$&        $-13.91_{0.30}^{0.36}$\\

\tableline
\end{tabular}
}
\end{center}
\end{table*}

\begin{table*}[h]
\begin{center}
\small
\caption{X-ray spectral parameters for 2 temperature fits.\label{tab:2t_fits}}
\setlength{\extrarowheight}{0.5em}
\begin{tabular}{rrrrrrrrrrrr}
\tableline\tableline
CXO Id.&    Counts&     Class&   Sp.Type&       n.p&                            nH&                           kT1&                           kT2&                   norm1/norm2\\
\tableline
                 797&    1465.5&      memb&O6IV+O9III&      59.6&          $1.27_{0.06}^{0.06}$&          $0.59_{0.04}^{0.04}$&                $5.4_{2.6}^{}$&                          40.1\\
                1168&    1955.7&        fg&          &      45.0&          $0.24_{0.05}^{0.05}$&          $0.37_{0.02}^{0.02}$&             $2.0_{0.2}^{0.3}$&                           2.3\\
         2197$^\dag$&   20978.5&      memb&O7I+O6I+O9&       0.0&          $1.30_{0.02}^{0.02}$&          $0.63_{0.02}^{0.02}$&             $3.1_{0.2}^{0.3}$&                           8.6\\
                2547&    7474.3&      memb&   O7.5III&      17.7&          $1.55_{0.04}^{0.04}$&          $0.74_{0.04}^{0.05}$&             $3.0_{0.3}^{0.4}$&                           3.8\\
         2926$^\dag$&   29905.6&      memb&   B3.5Ia+&       0.0&          $1.80_{0.02}^{0.02}$&          $0.71_{0.02}^{0.02}$&             $2.9_{0.2}^{0.2}$&                           6.6\\
                3895&    3194.5&      memb&      O8II&      27.8&          $1.60_{0.05}^{0.05}$&          $0.75_{0.05}^{0.06}$&             $3.4_{0.7}^{2.0}$&                          11.6\\
         4377$^\dag$&   15863.8&      memb&O5I+O3.5II&       0.0&          $1.50_{0.03}^{0.03}$&          $0.64_{0.03}^{0.03}$&             $6.1_{0.6}^{0.7}$&                           4.7\\
         4607$^\dag$&   50099.8&      memb&O6I+O5.5II&       0.0&          $1.06_{0.02}^{0.02}$&          $0.72_{0.02}^{0.02}$&             $3.2_{0.1}^{0.1}$&                           2.1\\
                5061&    7353.7&      memb&     O5.5V&       8.1&          $1.26_{0.05}^{0.06}$&          $0.68_{0.05}^{0.03}$&             $2.6_{0.2}^{0.3}$&                           2.1\\
                6534&     522.0&        fg&          &      95.2&              $0.00_{}^{0.03}$&          $0.84_{0.06}^{0.07}$&             $3.5_{1.0}^{2.7}$&                           1.9\\
                7883&    1823.9&      memb& WC6+O8III&       3.5&          $1.68_{0.13}^{0.12}$&          $0.49_{0.07}^{0.08}$&             $3.3_{0.5}^{0.6}$&                           8.8\\

\tableline
\end{tabular}
\tablecomments{$^\dag$: sources whose X-ray spectra suffers from significant pile-up. Spectral fits are statistically not acceptable and best-fit parameters uncertain.}
\end{center}
\end{table*}

%


\section{Coronal abundances}
\label{sect:abund}

Throughout our spectral analysis, we have consistently adopted the
plasma abundances estimated by \citet{mag07} for the brightest coronal
sources in the Orion Nebula Cluster (ONC). These were obtained from ACIS CCD
resolution spectra obtained with the $\sim$850\,ks {\em Chandra
Orion Ultra-deep Project} (COUP) observations. The spectra of a sample of
35 bright high signal-to-noise ratio X-ray sources ($>$10000 counts) were fit
individually with the {\sc vapec} model, i.e.\ the same optically-thin
thermal plasma emission model we have adopted here, but treating the abundances of individual elements as free fit parameters. Even with high S/N, CCD-resolution spectra only provide loose
constraints on most abundances. For each element, \citet{mag07} thus took
the median values of the sample as representative of the true
abundances.

We have attempted to estimate plasma abundances for our Cyg\,OB2 coronal sources or, at least, verify that they are compatible with the assumed ones. With respect to the ONC case discussed by \citet{mag07}, our Cyg\,OB2 spectra have significantly lower S/N, and, because of the larger absorption of the region with respect to the ONC, they specifically lack signal at low energies, where the prominent lines of elements such as N and O are found. On the other hand, we can rely on a much larger stellar sample. For these reasons, instead of fitting  individual spectra and taking the median abundances as in \citet{mag07}, we decided to investigate two alternative approaches. 

Our first approach consists of fitting stacked spectra of large samples of  Cyg\,OB2 members with either a single absorbed isothermal spectrum, or an absorbed two-temperature thermal spectrum. In this way, we hope to obtain average values of all parameters, including $N_H$, $kT$ (or $kT$s), and all the element abundances we can realistically constrain. One possible disadvantage of using stacked spectra is that biases that are hard to control might be introduced by the mixing of spectra with different plasma temperatures and subject to different absorptions. In particular, when using stacked spectra, we should be careful in interpreting the abundances of elements with prominent lines in the 1-2\,keV energy range, which is significantly affected by absorption for $N_H\ga 21.0$\,cm$^{-2}$, such as Ne and Mg.

Our second approach is to perform simultaneous fits of the CCD spectra of selected samples of Cyg\,OB2 X-ray sources. Each source within the sample is allowed to have its own absorption ($N_H$) and temperature ($kT$) but, for each element, the model abundance in the parameter estimation scheme is tied as a single parameter across all the difference sources being modeled. These simultaneous fits, performed using the {\sc vapec}  model within XSPEC, are computationally intensive and a simultaneous fit of all X-ray detected members is not practical. We instead divided the full sample according to ranges of net detected counts and performed simultaneous fits of the spectra in each subsample. Each of these spectral fits is thus independent and can be used to cross-check the results from other subsamples. 

In order to validate and tune these two methods, and to investigate possible biases, we have first applied the same procedures to the ONC stars using the COUP data. 

\subsection{Stacked spectra}

We have first considered a representative X-ray spectrum of COUP coronal sources, constructed by stacking the spectra of a sample of sources with reasonable quality X-ray spectra. Starting from the sample of COUP sources for which estimates of photospheric effective temperatures were available, we selected coronal sources by requiring $T_{eff}<10^4$\,K, thus excluding early type stars whose X-ray emission is not coronal. We then selected the X-ray sources whose spectral fits by \citet{get05} required two thermal components and had {\em low} best-fit values of the absorbing column density ($\log N_H<21.5$\,cm$^{-2}$). The stacked spectrum of the resulting 218 COUP sources contains $\sim 1.1$ million counts. We fit the spectrum with 1T, 2T, and 3T models, in all cases with free abundances for the same elements whose abundances were constrained by \citet{mag07}.  These were: O, Ne, Mg, Si, S, Ar, Ca, Fe and Ni.

\begin{figure}[!t!]
\epsscale{1.2}
\plotone{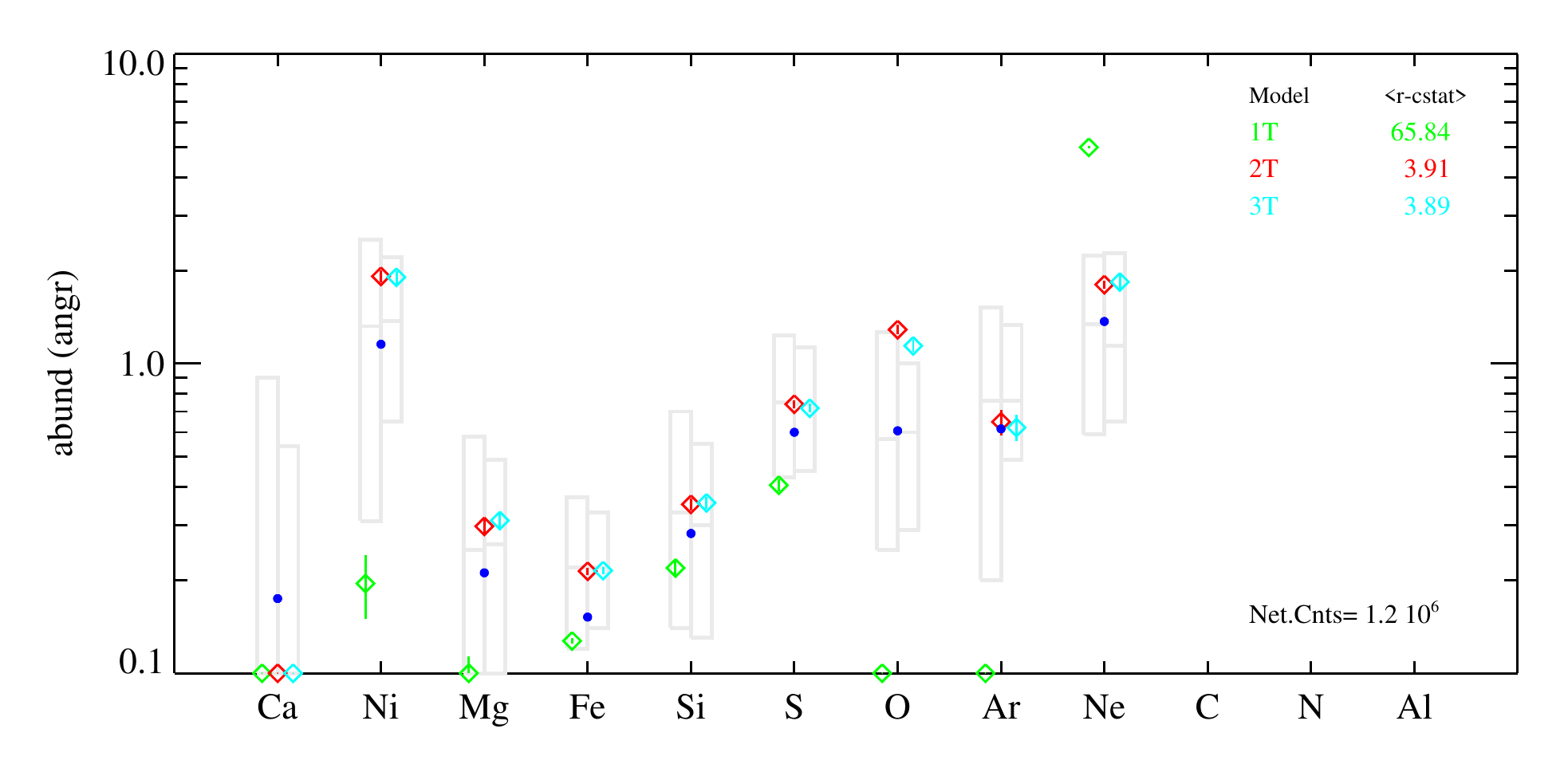}
\caption{Abundances derived from a stacked spectrum of a sample of COUP sources, namely those with estimates of photospheric effective temperature and mass, with effective temperature less than 10$^4$\,K, and with $\log N_H$ (cm$^{-2}$), from \citet{get05}, less than 21.5. Green, red, and cyan points, with 1$\sigma$ error bars indicate best-fit abundances obtained from fits with 1, 2, and 3 thermal components, respectively.  Blue dots indicate the weighted mean of abundances from the simultaneous fits shown in Fig.\,\ref{fig:COUP_simfit_abund}. Grey boxes indicate the median and 68\% interval of the abundances obtained by \citet{mag07} for stars in their ``low-absorption sample" (boxes on the left) and their ``count-limited subsample" (boxes on the right)  \label{fig:COUP_stacked_abund}}
\end{figure}

Results are shown in Fig.~\ref{fig:COUP_stacked_abund}: 2T and 3T models yield fits with very similar quality (reduced c-stat $\sim$3.9) and abundances roughly reproducing the \citet{mag07} results with some potentially interesting differences. Single-temperature fits, instead, are decidedly worse statistically and result in significant biases in the best-fit abundances. The most striking difference with respect to the \citet{mag07} abundances is for oxygen. We will see below that the determination of the oxygen abundance depends heavily on source absorption and we speculate that the value we derive from a stacked spectrum, the sum of individual spectra subject to  different $N_H$ values, may be biased. 

%
%

\begin{figure*}[]
\includegraphics[scale=0.62]{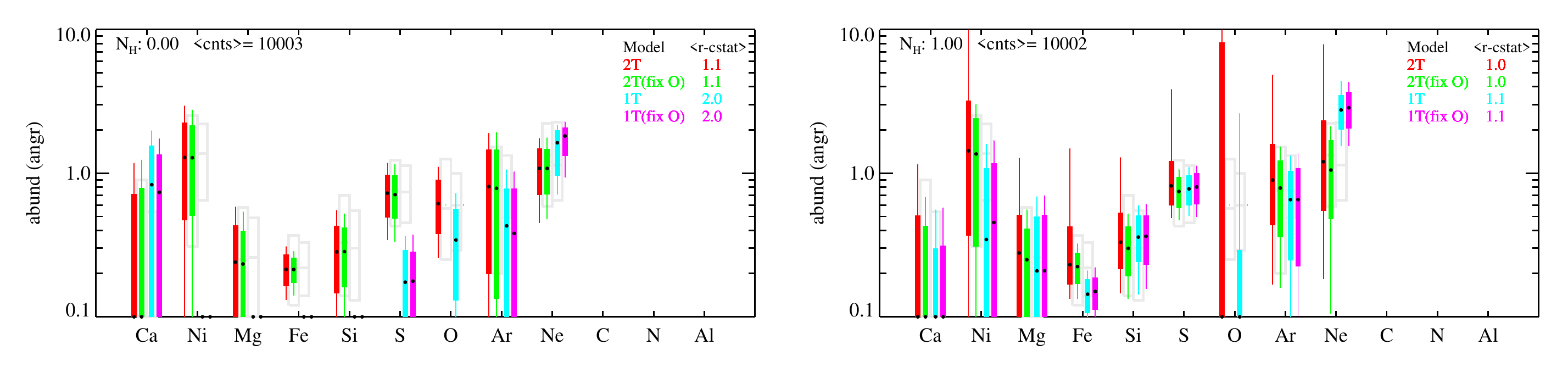}\\
\includegraphics[scale=0.62]{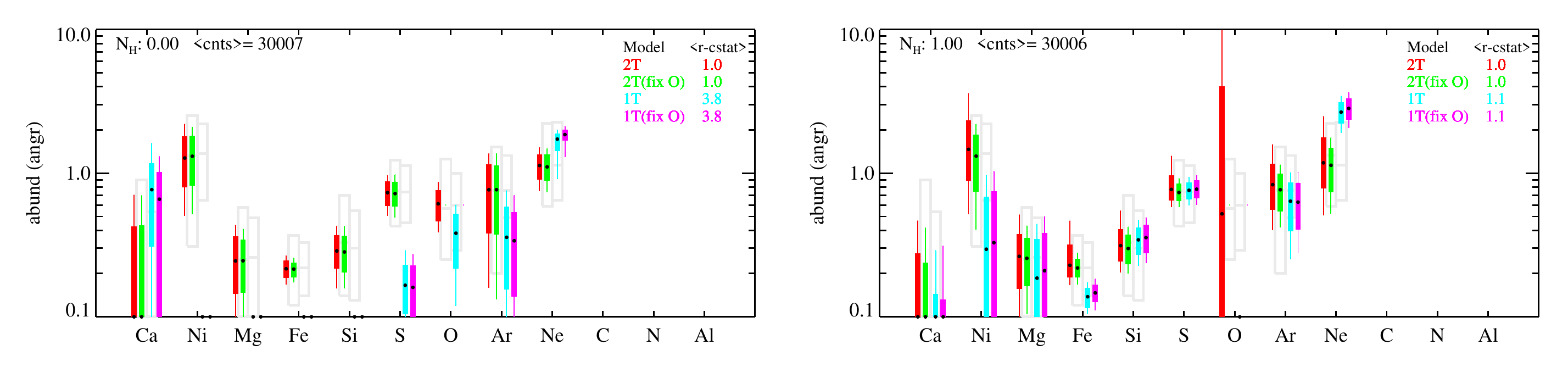}\\
\includegraphics[scale=0.62]{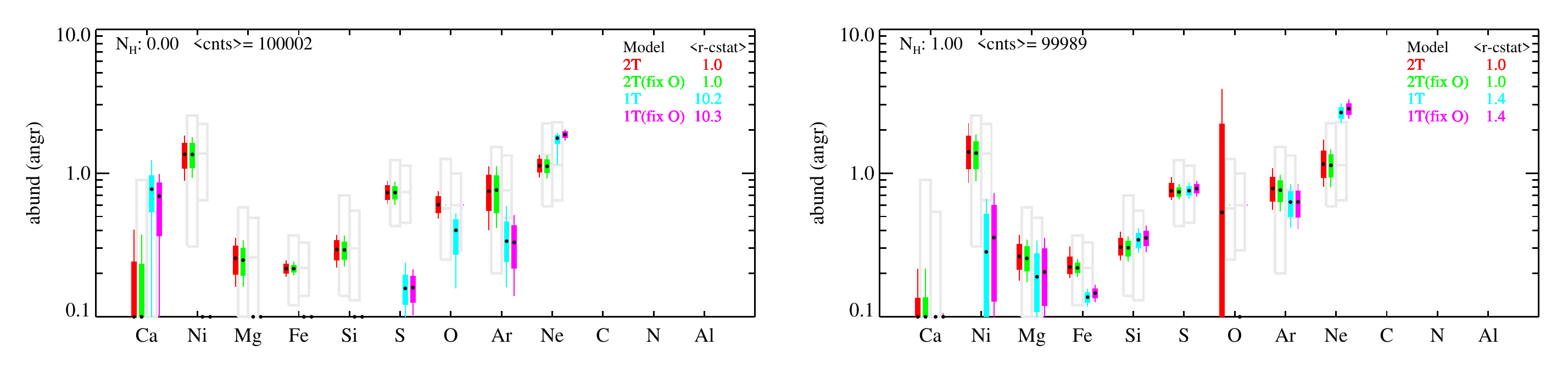}
\caption{Results of simulations to assess the ability to retrieve element abundances as a function of spectrum absorption ($N_H$) and  statistics (net counts). See text for details. Each panel refers to a different value of $N_H$, 0.0\,cm$^{-2}$ in the left-hand column and 1.3$\times 10^{22}$\,cm$^{-2}$ in the right-hand column, respectively, and net counts, 1, 3, and 10$\times10^4$, in the top, center, and bottom rows, respectively. Within each panel, the thick (thin) vertical bars show the 1$\sigma$ (90\%) dispersions of the abundances derived for each fitted elements for a set of 1000 simulations. Median values are indicated by small black dots within the bars. Four estimates are shown, each derived from fitting the simulated spectra with a different model: a 2T model with all shown abundances treated as a free parameter (red bars), a 2T model in which Oxygen was fixed at the values derived by \citet{mag07}, i.e. 0.6 times the solar {\sc angr} abundances (in green), a 1T model (cyan) and a 1T model with fixed Oxygen abundance (magenta). For each spectral model, the average reduced c-stat for fits to the simulated spectra is given in the upper-right corner, while the input $N_H$ and the average number of net counts in the simulated spectra are given in the upper-left corner. \label{fig:COUP_abund_sim}}
\end{figure*}

Next, we have tried to assess, through Monte Carlo simulations, the effects of photon statistics and absorption on the determination of abundances, so as to help in the interpretation of our Cygnus~OB2 results. Our simulations were based on the 2T model fit to the ONC data discussed above, but with abundances fixed to those of \citet{mag07}. Assuming that low-mass stars in the ONC and Cygnus~OB2 share the same intrinsic X-ray spectrum, we verify our ability to retrieve the input abundances by generating simulated spectra with varying values of $N_H$ and different levels of average counting statistics ($10^4$, 3$\times10^4$, and $10^5$ net counts) covering the range of our Cygnus~OB2 stacked spectra (see below). For each pair of $N_H$ and net counts, we produce 1000 simulated spectra and fit each with both 1T and 2T models, initially assuming the solar {\sc angr} abundances from \citet{and89} and fitting the same elements constrained by \citet{mag07}. Moreover, we performed additional 1T and 2T fits keeping the Oxygen abundance fixed at the \citet{mag07} value (0.6$\times$solar). Figure  \ref{fig:COUP_abund_sim} shows the mean abundances and relative uncertainties that are expected for two values of $N_H$ and three values of net counts.

At low $N_H$, input abundances are retrieved when fitting the spectra with 2T models, while 1T models produce biased results, in qualitative agreement with the results on the real ONC stacked spectrum.
We are, however, most interested in the results for $N_H\sim 1.0 \times 10^{22}$\,cm$^{-2}$ (i.e.\ a representative value for Cygnus~OB2): $a)$ The abundance of oxygen cannot be constrained,  even with $10^5$ counts, and must therefore be fixed (at the \citet{mag07} level); $b)$ at least $\sim 3 \times 10^4$ counts are needed to constrain Fe and Ne; $c)$ 1T fits to the absorbed 2T models are statistically {\em acceptable} (i.e.\ average reduced c-stat = 1.1, 1.1, and 1.4 at $10^4$, $3 \times 10^4$ and $10^5$ counts),  but produce biased  abundances (e.g.\ low Fe and high Ne). 

\begin{figure*}[!t!]
\epsscale{1.17}
\plottwo{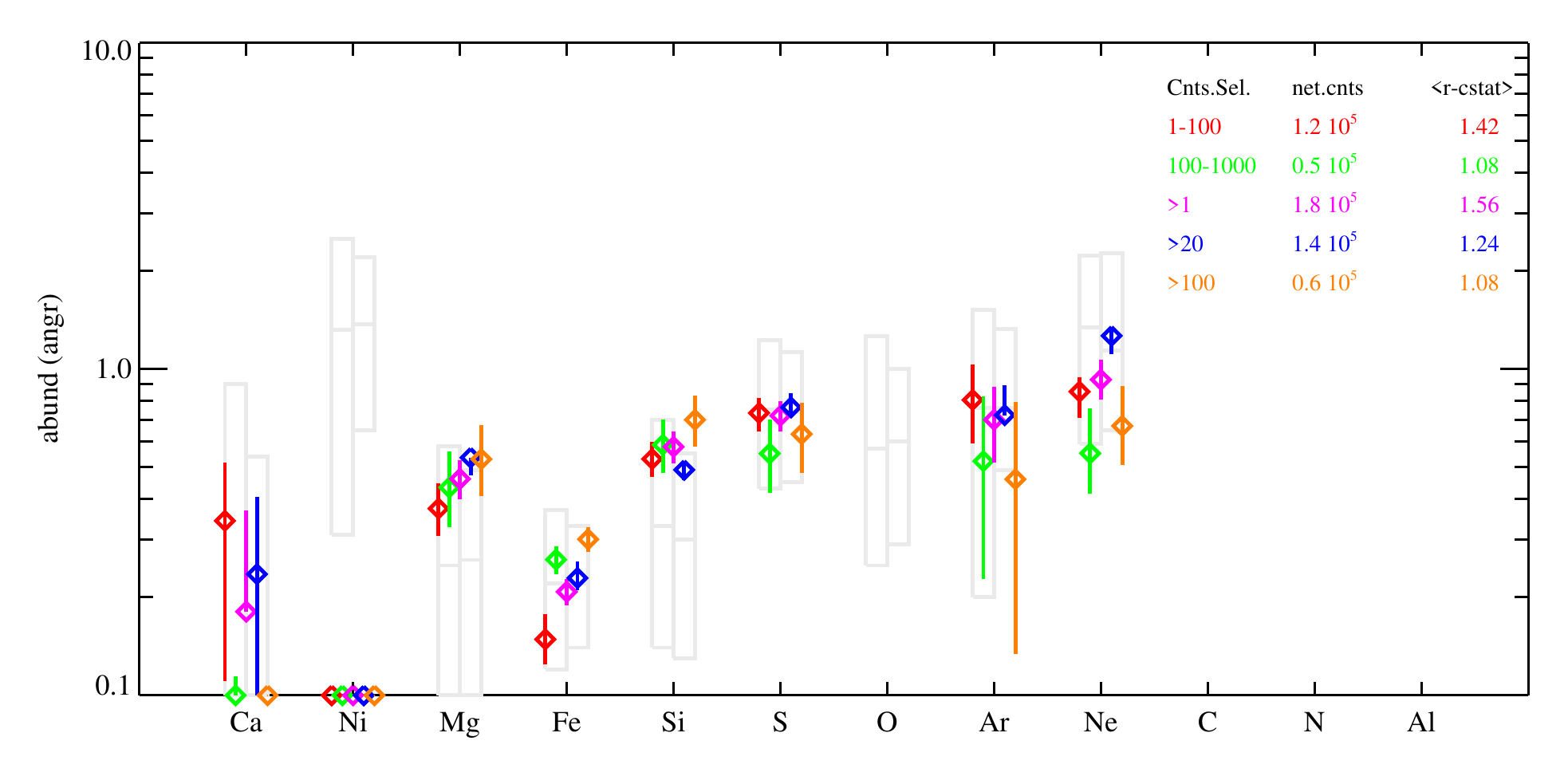}{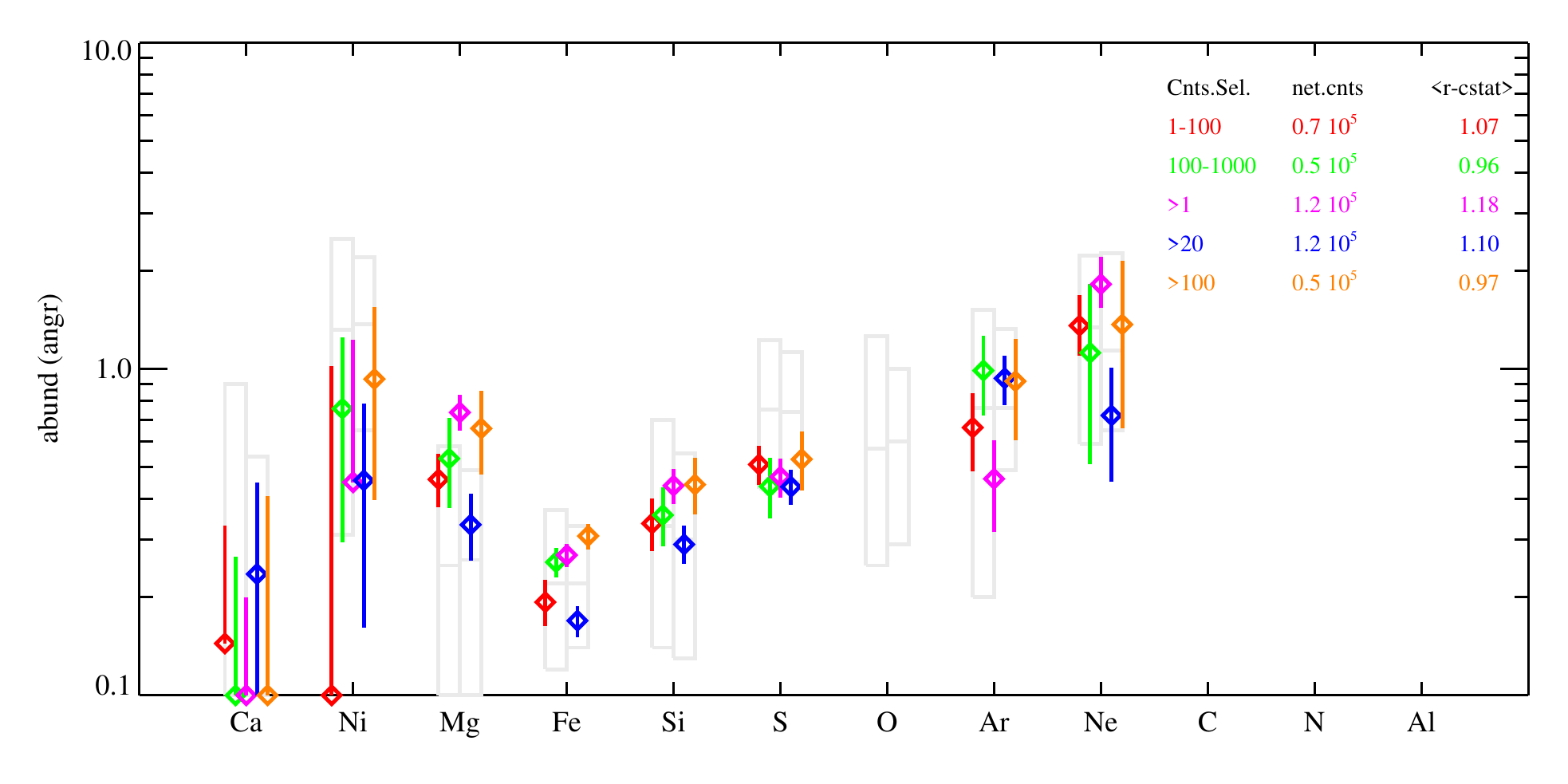}
\caption{Abundances derived from a stacked spectrum for five samples of Cyg\,OB2 members selected in five different ranges of net-counts, as indicated in the legend in the upper-right corner of each panel. The left-hand panel refers to the full samples of members, the one on the right to samples of members for which the 1T spectral fits yielded {\em reasonable} parameters ($0.5<N_H<5.0\times10^{22}$\,cm$^{-2}$ and $kT>1.0$\,keV). 
In all cases, stacked spectra were fit with a 2T model with the Oxygen abundance fixed to the \citet{mag07} values (0.6$\times$ solar, in the {\sc angr} table). Note that in a small number of cases upper or lower error uncertainties are not plotted as their estimation within XSPEC failed.  
\label{fig:CygOB2_stacked_abund}}
\end{figure*}

On the basis of the above results, we present in Fig.\,\ref{fig:CygOB2_stacked_abund} 2T fits to the stacked Cyg\,OB2 spectra, with O fixed to 0.6$\times$solar ({\sc angr}). The panel on the left shows abundances derived from stacked spectra constructed for five samples: all members with between 1 and 100 net counts, those with 100 to  1000 net counts, and those with more than 1, 20, and 100 net counts. Very similar results are obtained, with abundances mainly consistent with those of \citet{mag07}, but with some differences in Mg, Si, and Ni (higher Mg and Si, and lower Ni). In order to assess the effect of outliers, e.g. misclassified non-members or members with peculiar  X-ray spectra, we also considered samples with more homogeneous X-ray spectral characteristics, namely those for  which the 1T spectral fits yielded $0.5<N_H<5.0\times10^{22}$\,cm$^{-2}$ and $kT>1.0$\,keV. Results are shown in  the right-hand panel of Fig.\,\ref{fig:CygOB2_stacked_abund}. The main differences we see with respect to the full samples is that in the restricted ones the Ni and Si abundances are now consistent with the COUP values.

\subsection{Simultaneous fits}

\begin{figure*}[h]
\epsscale{1.2}
\plotone{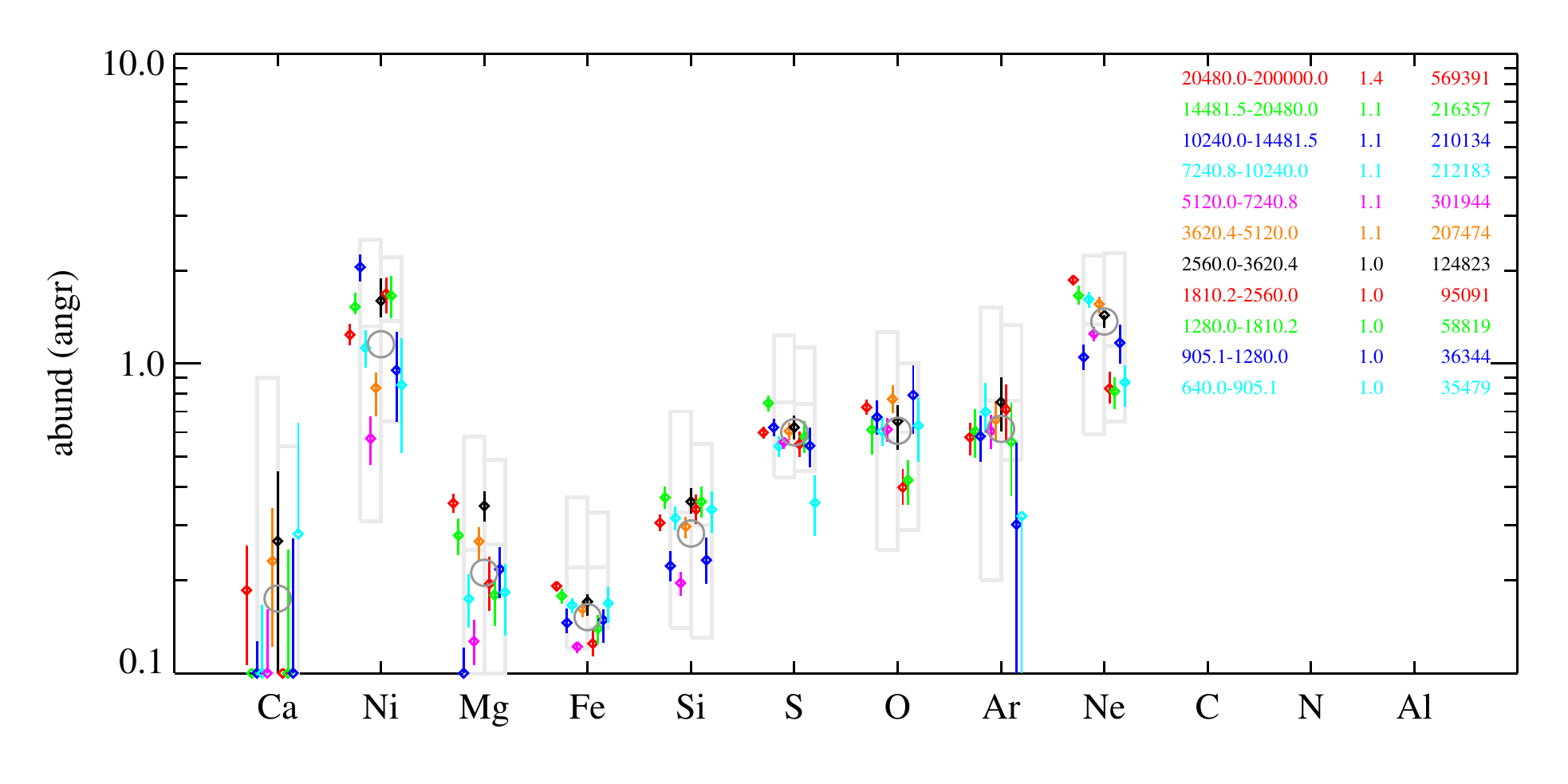}
\caption{Abundances derived from simultaneous fits of different and independent samples of ONC sources from the COUP survey. The samples are selected so to contain sources with net counts in selected ranges, plotted with decreasing counts from left to right, as indicated in the legend (note that some colors are repeated). The spectra were fit with 2T models, allowing $N_H$ and the two temperatures to vary for each source.  Reduced c-stat values of the fits and total number of counts in the spectra are indicated beside each count range. Large empty circles indicate the uncertainty-weighted mean of the abundances for all count ranges.  \label{fig:COUP_simfit_abund}}
\end{figure*}

Simulations to assess the effect of simultaneously fitting multiple spectra with 1T models, each with individual $N_H$ and kT, and with shared but variable abundances, are not easy to produce. Each simultaneous fit requires a large amount of computing time, especially as the number of sources to fit simultaneously increases. For COUP sources, where sources are fewer and brighter and the $N_H$ is lower, we performed simultaneous fits using 2T models,  for several independent source samples. Results, shown in Fig.\,\ref{fig:COUP_simfit_abund}, largely confirm the abundances of \citet{mag07}, and those derived from the fit of the stacked COUP spectra in Fig.~\ref{fig:COUP_stacked_abund}
(although we obtain marginally lower Fe abundances).

\begin{figure*}[t]
\epsscale{1.17}
\plottwo{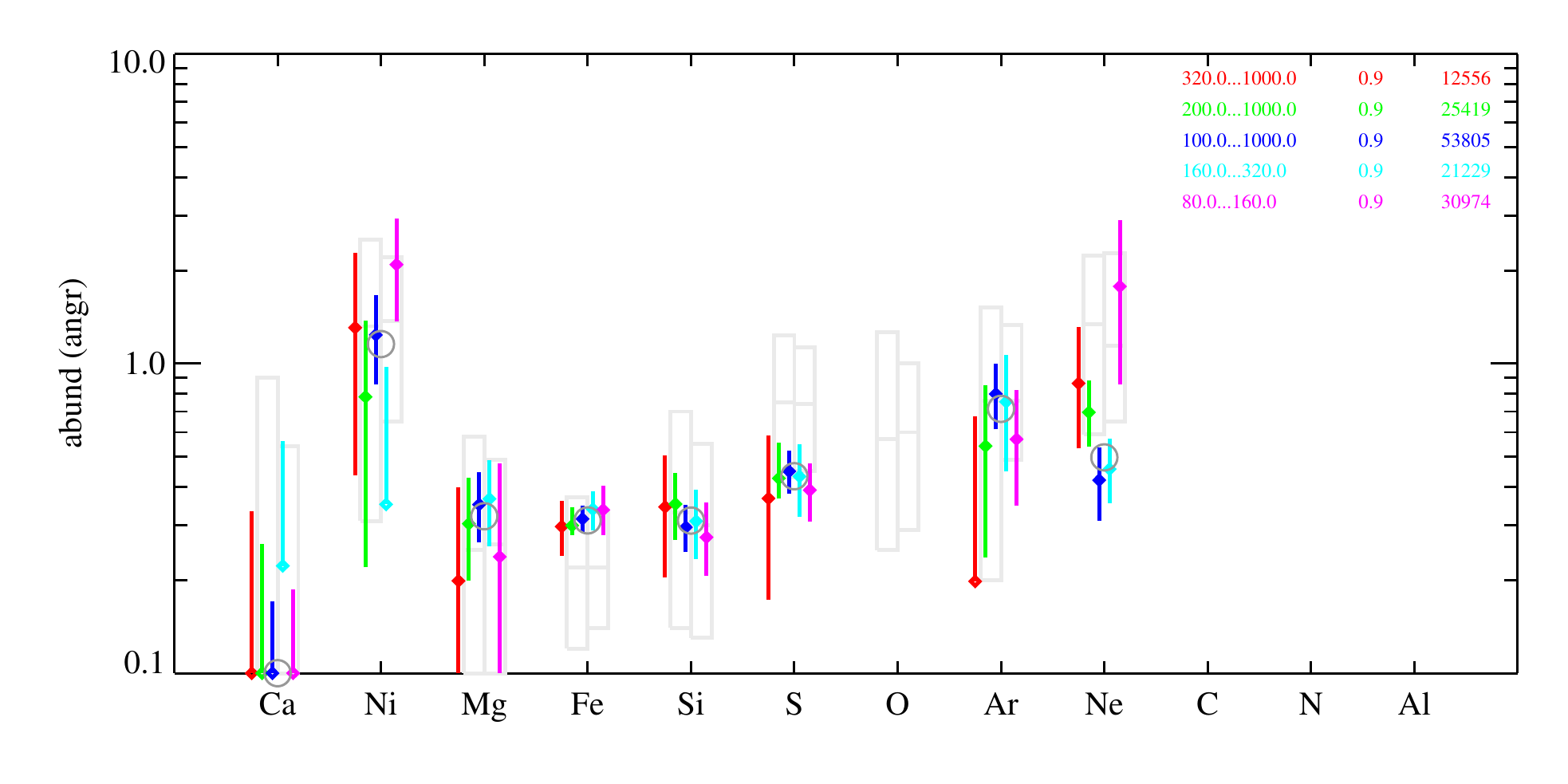}{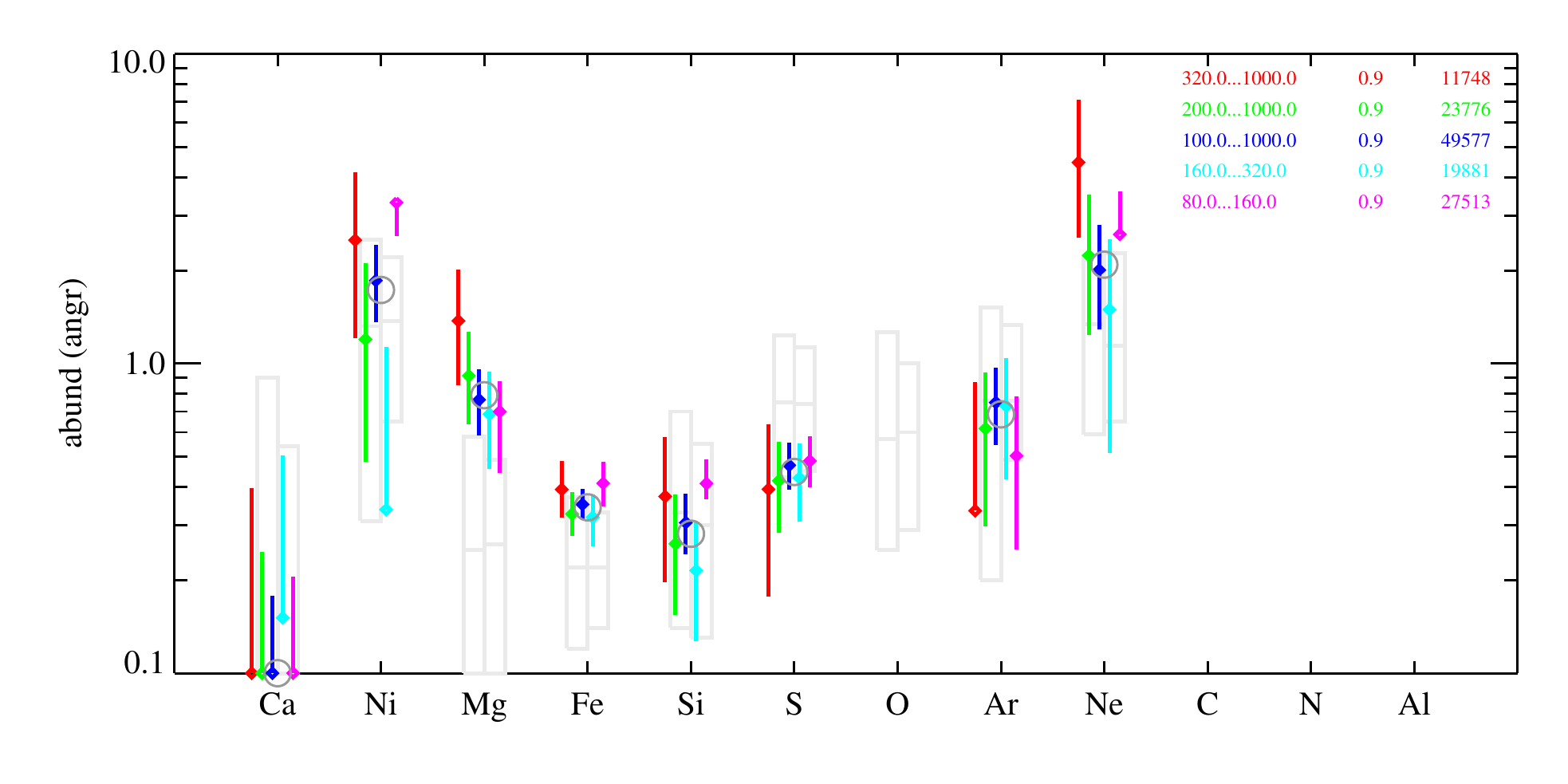}
\caption{Abundances derived from simultaneous fits for different (and not all independent) samples of Cygnus\,OB2 members.  In the left panel, samples are selected so as to contain all members detected in X-rays  with net counts in selected ranges, as indicated in the legend in the upper-right. Spectra were fit with 1T models, allowing $N_H$ and $kT$  to vary for each source.  Reduced c-stat values for the fits and the total number of counts in the spectra are indicated beside each count range. Gray circles indicate uncertainty weighted mean abundances computed from all values with fully defined uncertainties.  The right panel  is analogous to that on the left, but samples are restricted to X-ray sources whose individual 1T spectral fits yield $N_H$ and $kT$ values that are consistent with the  member population (more specifically, $0.5<N_H<5.0\times10^{22}$\,cm$^{-2}$ and $kT>1.0$\,keV).  \label{fig:CygOB2_simfit_abund}}
\end{figure*}

The results for Cygnus OB2 stars, as shown in Fig.\,\ref{fig:CygOB2_simfit_abund}, are harder to interpret. We are indeed forced to use 1T models (with fixed Oxygen). Had we used 2T models, in addition to further increasing the computation time, fit parameters would probably become largely unconstrained. We do not have a firm understanding of how the choice of 1T models for the simultaneous fits biases abundances. Looking at the right-hand panel in the figure (for the somewhat cleaned samples, excluding sources with peculiar spectra) we derive, with respect to \citet{mag07}, somewhat higher Fe and Mg (and lower S) abundances. We also notice similarities and some differences with respect to the stacked spectra analysis, the latter possibly due to the aforementioned biases. In particular, focusing on the ``cleaned samples'':

\begin{itemize}
\item Ne appears consistent  between the two analyses and with the ONC values (all analyses). 
\item the Fe abundances from stacked spectra and simultaneous fits are consistent, and only slightly larger than those of ONC stars, as estimated by both \citet{mag07} and our stacked spectra analysis. ONC values from simultaneous fits are instead slightly lower. 
\item the Mg abundances, by the two separate analyses, are consistent, and 2-3 times higher than in the ONC (all analyses).
\item the Ni abundances from the two approaches appear to be consistent and similar to the ONC values. 
\end{itemize}




\section{Discussion}
\label{sect:discussion}

We now make use of the results of the above analyses to discuss the properties of the coronal plasma in Cygnus~OB2 stars, in terms of temperature, flux, and abundances, and to investigate the relation between X-ray and optical  extinction toward the region.

\subsection{Coronal plasma}
\label{sect:disc_plasma}

At first glance, the coronal plasma properties of Cygnus~OB2 members differ somewhat from those of other young stars, e.g., those in the ONC, as determined by the COUP studies \citep{pre05}. In particular, we successfully fit the X-ray spectra of almost all the Cygnus~OB2 members with isothermal models, while most of the ONC stars required two thermal components. Moreover the mean plasma temperature of Cygnus~OB2 sources, $\sim$3\,keV, appears higher than the average temperature of COUP sources, e.g. $\sim$2.2\,keV for the emission-measure weighted temperature of low mass stars with $\log L_X>$30.0\,erg/s (or $\sim$2.4\,keV for $\log L_X>$30.5\,erg/s). 

As demonstrated in Appendix\,\ref{sect:ONCsim}, however, it is important to assess the biases introduced by fitting our spectra, which are most likely intrinsically multi-temperature, with a single thermal component, even if it is statistically sufficient to represent the observed spectra. This is due mainly to the high extinction, which obscures the low-energy part of the spectra, and, to a lesser extent, to the large distance of Cygnus\,OB2 sources resulting in spectra with a low signal-to-noise ratio. Appendix\,\ref{sect:ONCsim} describes our efforts to quantify these effects by assuming that our sources have intrinsic spectra similar to those of the closer and much less absorbed young stars in the Orion Nebula Cluster, as characterized by \citet{get05}, and in particular, to those whose spectrum was fit with two thermal components. Using these templates, we simulate the effect of higher extinction and larger distance on the spectral characterization of the sources.

We conclude that, if our Cygnus~OB2 sources have X-ray spectra that are intrinsically similar to those of their ONC counterparts: $i)$ our isothermal model fits are consistent with intrinsically 2-temperature spectra; $ii)$ our best-fit values of $N_H$ are underestimated, on average, by $\sim2\times 10^{21}$cm$^{-2}$ or, considering the average $N_H$ of Cygnus~OB2 stars, $\sim$20\%, $iii)$ our source fluxes are also underestimated by $\sim$0.11\,dex, $iv)$ the plasma temperatures we obtain are similar (0.1\,keV lower, on average) to those of the hot component of a two-temperature model (and $\sim$0.40keV higher than the average temperature). Overall, the Cygnus~OB2 X-ray spectra thus appear compatible with those of ONC stars. 

\begin{figure*}[t]
\epsscale{1.17}
\plottwo{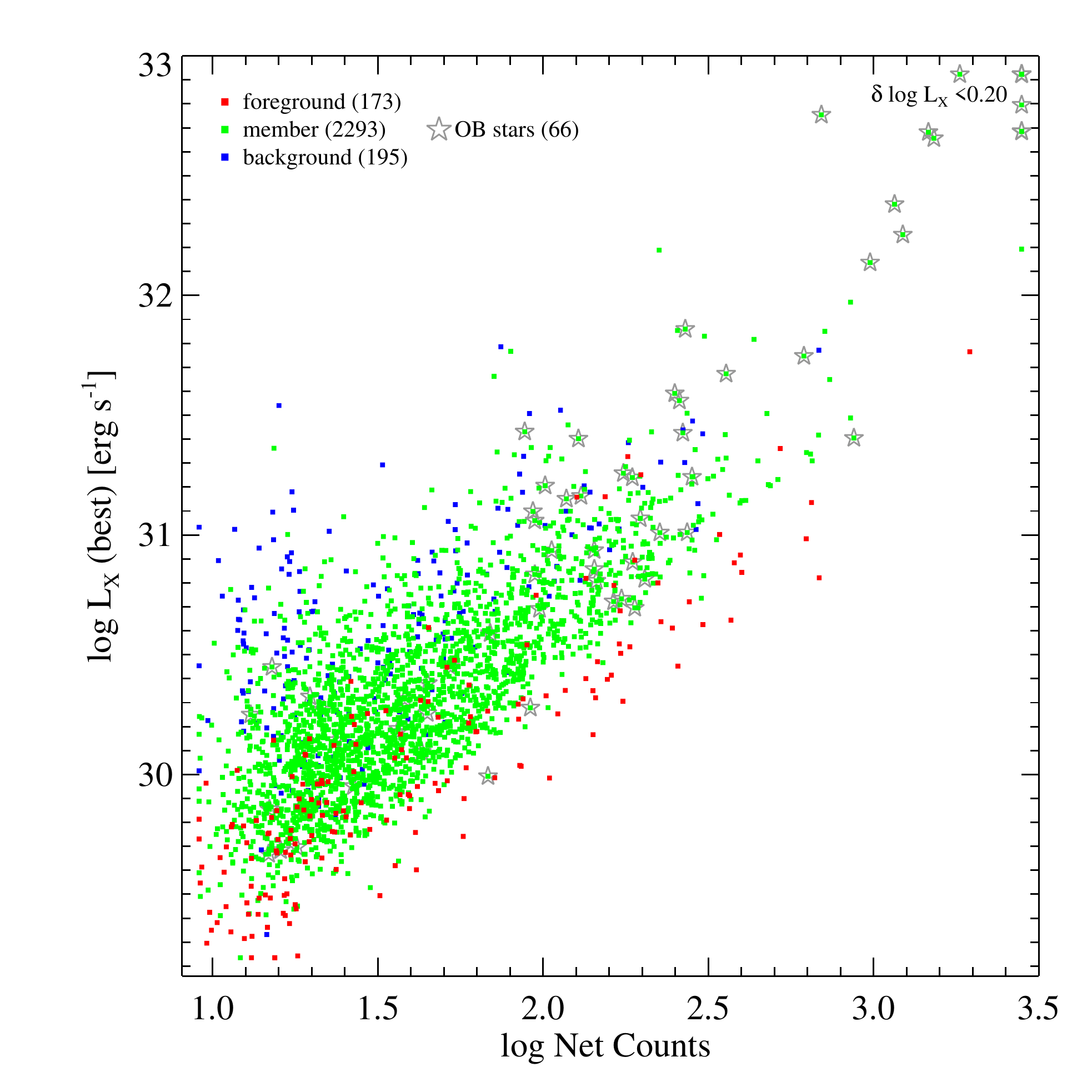}{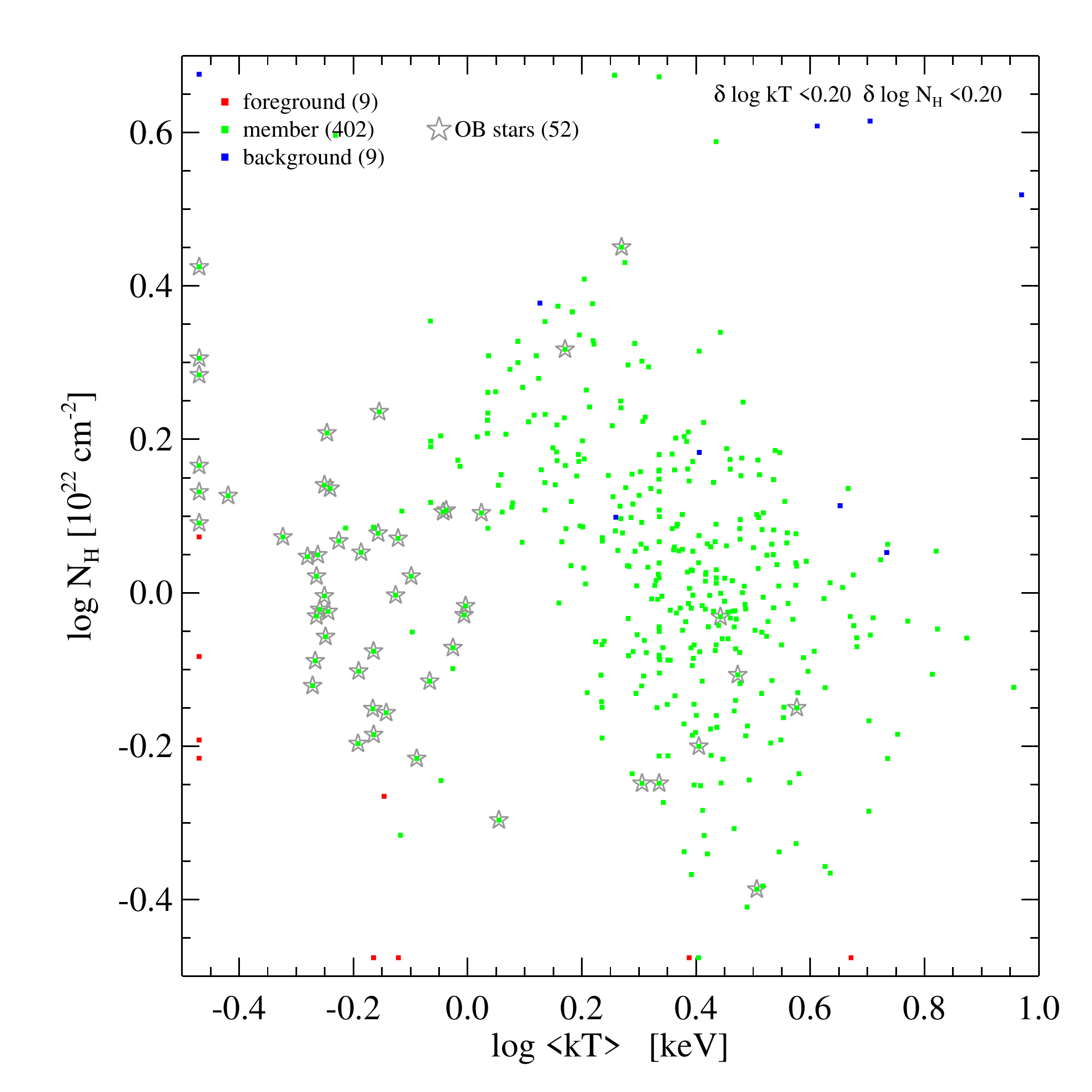}
\plottwo{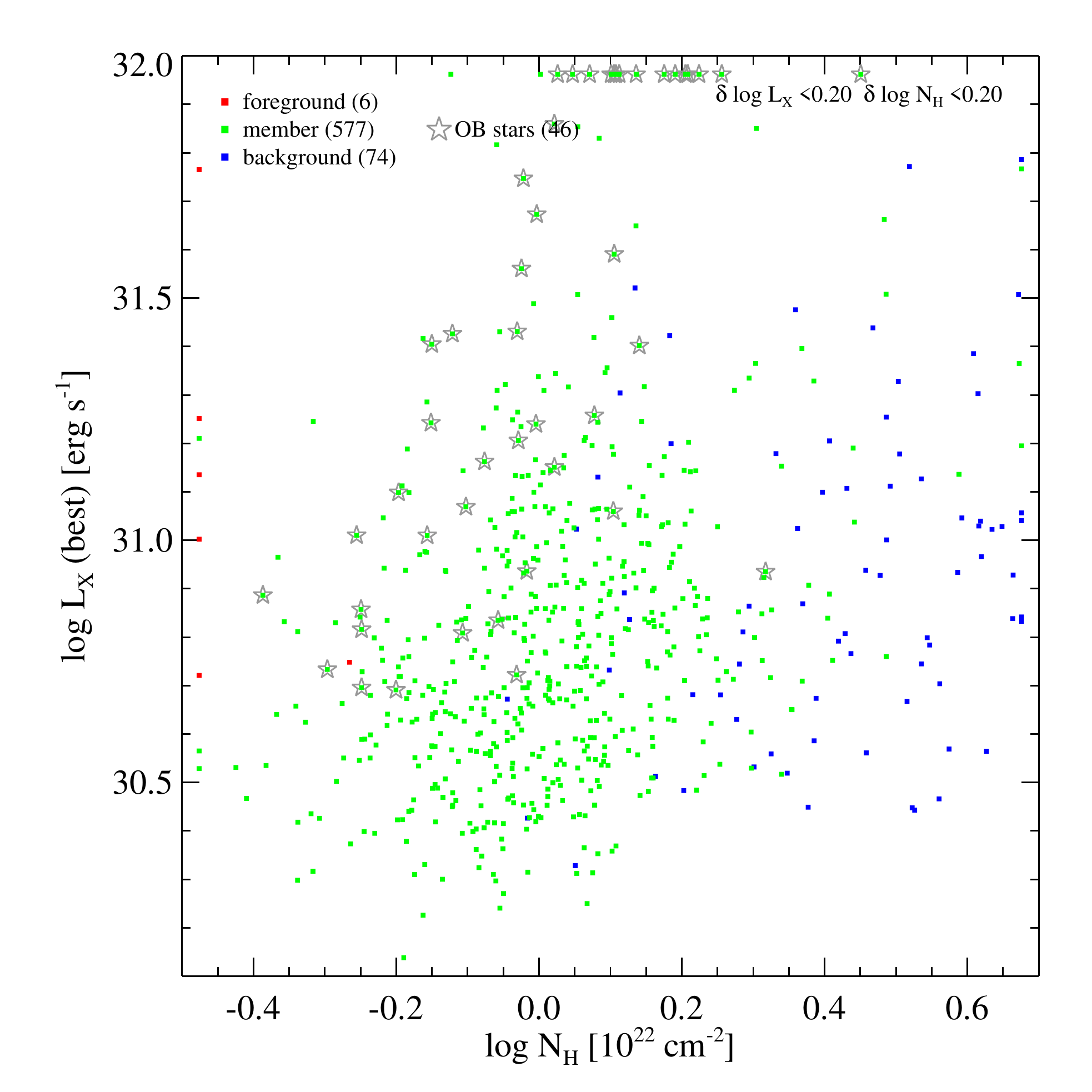}{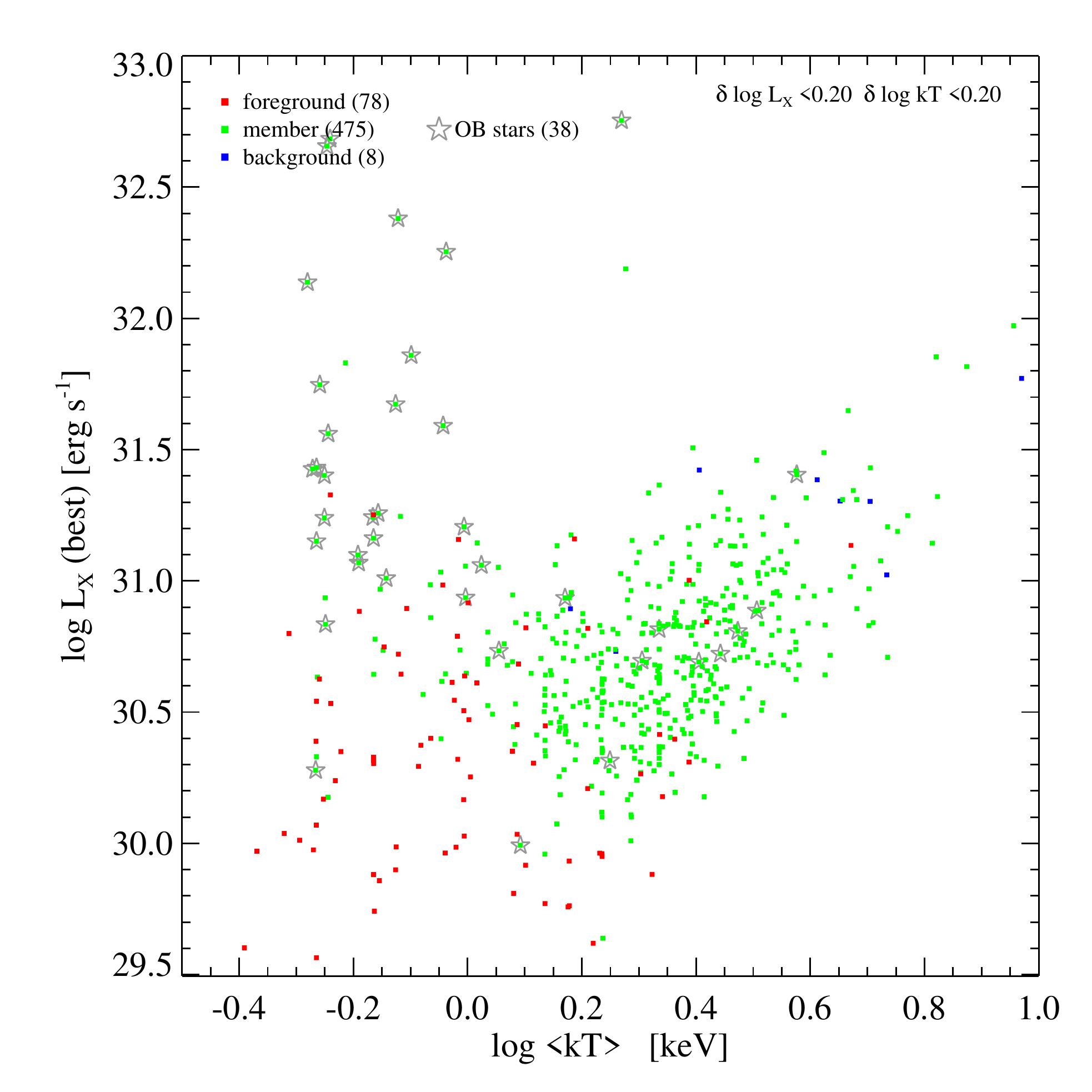}
\caption{[Upper-left] X-ray luminosities vs. net detected counts, separately for members, foreground, and background objects, as indicated in the legend in the upper-left corner. OB stars are marked with a star symbol. The sample is limited to sources with $1\sigma$ uncertainty on the X-ray luminosities  $<$0.2\,dex (as indicated in the upper-right corner). A handful of stars in the upper-right corner, with $L_X$ and/or Net Counts larger than the plot limits are plotted close to the axes. [Upper-right] same as above for $N_H$ vs. average $kT$. [Lower-left]  same as above, for $L_X$ vs. $N_H$.  [Lower-right] same as above for $L_X$ vs. average $kT$. \label{fig:LxCntsNHKT}}
\end{figure*}

Figure\,\ref{fig:LxCntsNHKT} shows the run of $L_X$ with detected net photons, the relation between $N_H$ and $kT$, and between $L_X$ and these latter two parameters. Only sources with relatively well-determined parameters are plotted, i.e. with 1$\sigma$ uncertainties on the plotted quantities $<$0.2dex. Focusing on low-mass Cygnus~OB2 members, i.e. excluding foreground and background objects, as well as OB members, we notice several points. First, $N_H$ and $kT$ are anti-correlated. This is most likely a spurious effect due to the well known anti-correlated shape of the $\chi^2$/c-stat space in the two fit-parameters. It is also fully confirmed by our simulations of ONC sources performed at fixed input $N_H$ showing the exact same trend.  $L_X$ and $N_H$ appear mostly uncorrelated, although a dearth of high-$L_X$, low-$N_H$ stars may be observed. 
Again, our simulations of ONC sources with fixed input $N_H$ reproduce this trend, which can thus be considered a spurious effect due to the statistical uncertainties on $N_H$ and the expected correlation between these uncertainties and the uncertainties on the absorption correction for the flux, and thus on $L_X$.
Finally, $L_X$ and $kT$ appear directly correlated. Our simulations of ONC sources instead produce an inverse correlation, which is easily understood in terms of the spurious $N_H$-$kT$ and $L_X$-$N_H$ (anti)correlations discussed above. 
The observed direct correlation can thus be considered even more significant and physical. This is in qualitative agreement with previous results on the correlation between coronal X-ray flux and plasma temperature \citep[e.g.][]{per04,pre05,joh15a}. 



\subsection{Coronal abundances}
\label{sect:disc_abund}

The abundances of elements in stellar coronae are known to differ from those in the underlying photosphere \citep[e.g.][]{dra96,dra02,lam15}.  While the exact mechanism, or mechanisms, responsible for the chemical fractionation have not been securely identified, the differences appear to depend both on magnetic activity level and on spectral type \citep{dra03a,tel07a,lam15,woo18}. Our Cygnus~OB2 sample of young low-mass stars represents a potentially valuable set of observations with which to investigate chemical fractionation further.

Overall, our abundance results for the T~Tauri stars in Cygnus OB2, presented in \S\,\ref{sect:abund}, appear to show characteristics of the ``inverse First Ionization Potential (FIP) Effect'' (iFIP), in which elements with low FIP (FIP$ \leq 10$~eV, e.g., Si, Mg, Fe) are depleted relative to elements with high first ionization potentials (FIP$ \geq 10$~eV, e.g., N, Ne, Ar).  This abundance pattern is opposite to that characterizing the solar corona, in which low FIP elements are enhanced \citep[e.g.][]{lam95}.  While hints of iFIP abundance patterns were present in early CCD resolution spectra of active stars obtained by the Advanced Satellite for Cosmology and Astrophysics ({\it ASCA}; see, e.g., \citealt{dra96} for a review), high resolution spectra obtained by {\it Chandra} and {\it XMM-Newton} provided the first definitive and detailed iFIP abundance information \citep{bri01,dra01}.  
Empirically, for both PMS stars and main sequence stars with little or moderate magnetic activity, the presence and strength of a solar-like FIP effect or an iFIP effect depends primarily on the spectral type \citep{tel07,gud07,lam15,woo18}. The most active stars, however, seem to depart from this relationship and to show strong iFIP abundance patterns \citep{woo18}. While the mechanism underlying the chemical fractionation responsible for FIP-based effects has not been firmly identified, the most promising explanation is based on the ponderomotive forces experienced by ions in the chromosphere \citep[e.g.][]{lam09}.


The abundance analysis of {\it Chandra} observations of young stars in the ONC by \citet{mag07}, and now also in Cygnus~OB2, firmly establish T~Tauri stars in the iFIP category, in keeping with expectations based on their ``saturated'' levels of magnetic activity. The dependence of FIP bias on spectral type \citep{woo18} cannot easily be investigated with our dataset. We, however, note that the Fe abundances and Fe/Ne ratios from our stacked spectra analysis (Fig.\,\ref{fig:CygOB2_stacked_abund}) for stars with more or less than 100 detected counts (and thus, on average, different $L_X$, cf. Fig.\,\ref{fig:LxCntsNHKT}) are both consistent with the trends of higher Fe and lower Fe/Ne for earlier-type (and X-ray brighter) stars \citep{tel07,gud07,woo18}.

\subsection{X-ray vs. optical extinction}
\label{sect:NHAVratio}

We used our extensive dataset to investigate the relation between optical  extinction, $A_V$, and X-ray absorption, as parametrized by the hydrogen column density $N_H$. The $N_H/A_V$ ratio is a measure of the gas-to-dust ratio of the interstellar medium (ISM), in our case toward and in the environment of our Cygnus\,OB2 stars. While $A_V$ depends on the content and properties of dust in the ISM, the X-ray absorption  depends on the column density of heavy elements which, for a given chemical composition of the ISM, is proportional to $N_H$. Different techniques and spectral bands (UV, radio, and X-rays) have been used to estimate the $N_H$ column density and the optical extinction toward a variety of objects---stars of different kinds and in different environments,  supernova remnants, X-ray binaries---see, e.g.\ \citet{zhu17} for a summary. The precise value of the $N_H/A_V$ ratio, and its eventual dependence on spatial and/or physical parameters, is debated. Several slightly different ratios have been determined even when using the same objects and spectral bands, e.g.\ 1.8$\times 10^{21}$\,cm$^{-2}$/mag and 2.2$\times 10^{21}$\,cm$^{-2}$/mag by \citet{pre95} and \citet{ryt96}, respectively.

A few determinations have specifically targeted star forming regions, often finding somewhat lower $N_H/A_V$ ratios with respect to the ``galactic values'' referenced above. \citet{vuo03}, for the highly absorbed X-ray spectra of stars in the $\rho$~Ophiuchi region, found $N_H/A_V=1.6\times 10^{21}$\,cm$^{-2}$/mag when adopting ``standard'' ISM abundances and transforming the J-band extinction to $A_V$ using a standard extinction law with $R_V=3.1$.  This $N_H/A_V$ ratio is somewhat lower than most estimates and has since been adopted in several works \citep[e.g.][]{get11,ski17}.

The above derivation was obtained from a sample of relatively absorbed stars, most with $N_H\gtrsim 10^{22}$\,cm$^{-2}$ (median $\sim 2\times10^{22}$\,cm$^{-2}$). Indeed, because of the large relative uncertainties in the determination of $N_H$ from X-ray spectral fits, especially for small $N_H$ values, the relation is more easily investigated using stars in more absorbed regions, just like $\rho$ Ophiuchi and Cygnus\,OB2. 
More recently, \citet{has16} have investigated the matter using $N_H$ values obtained by the COUP collaboration based on fitting X-ray spectra of stars in the ONC \citep{get05}, and  optical extinction values obtained from spectral types, nIR colors, and a standard extinction law. An $N_H/A_V$ ratio of 1.4$\times 10^{21}$\,cm$^{-2}$/mag was obtained, which is even lower than that of \citet{vuo03}.

\begin{figure*}[!t!]
\epsscale{1.17}
\plottwo{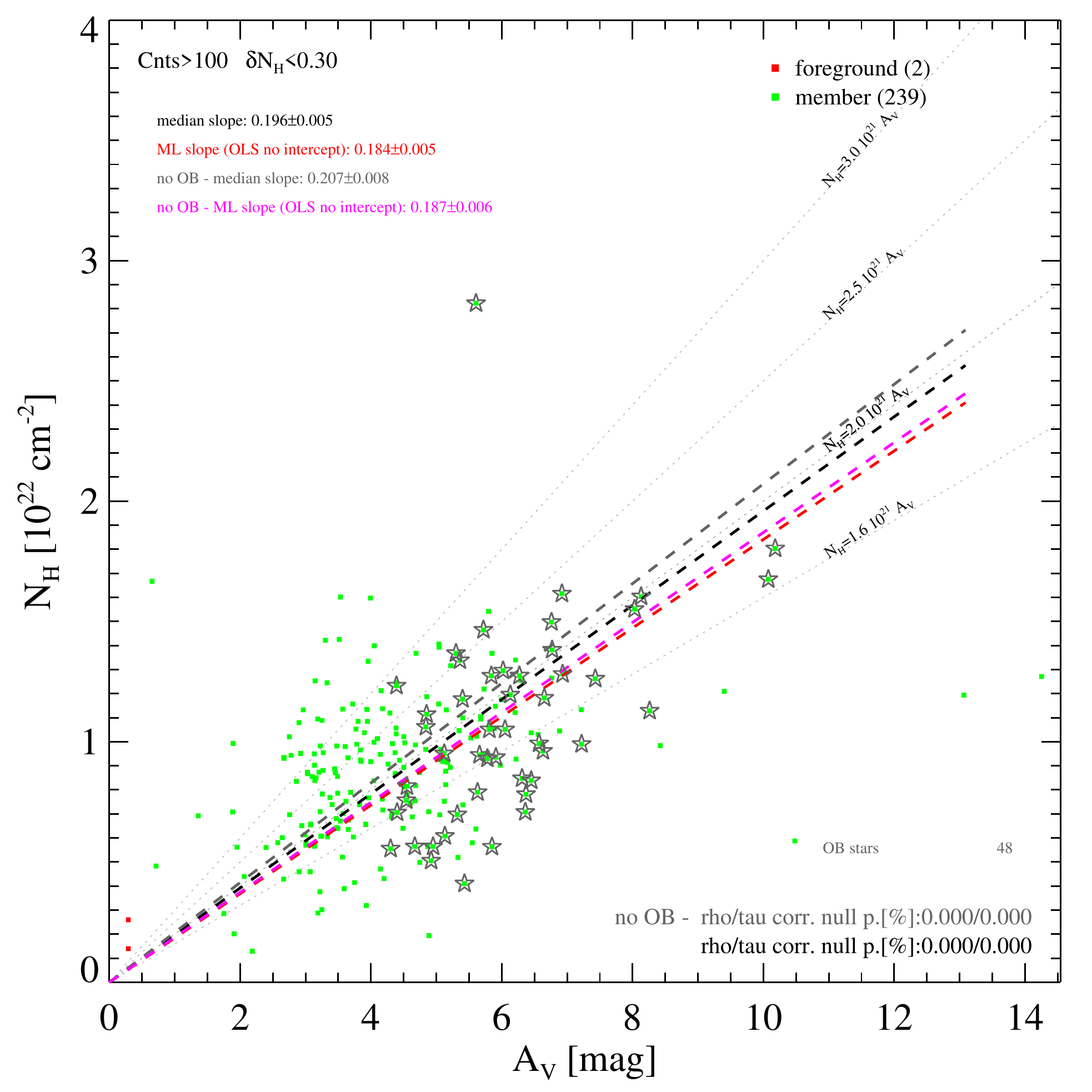}{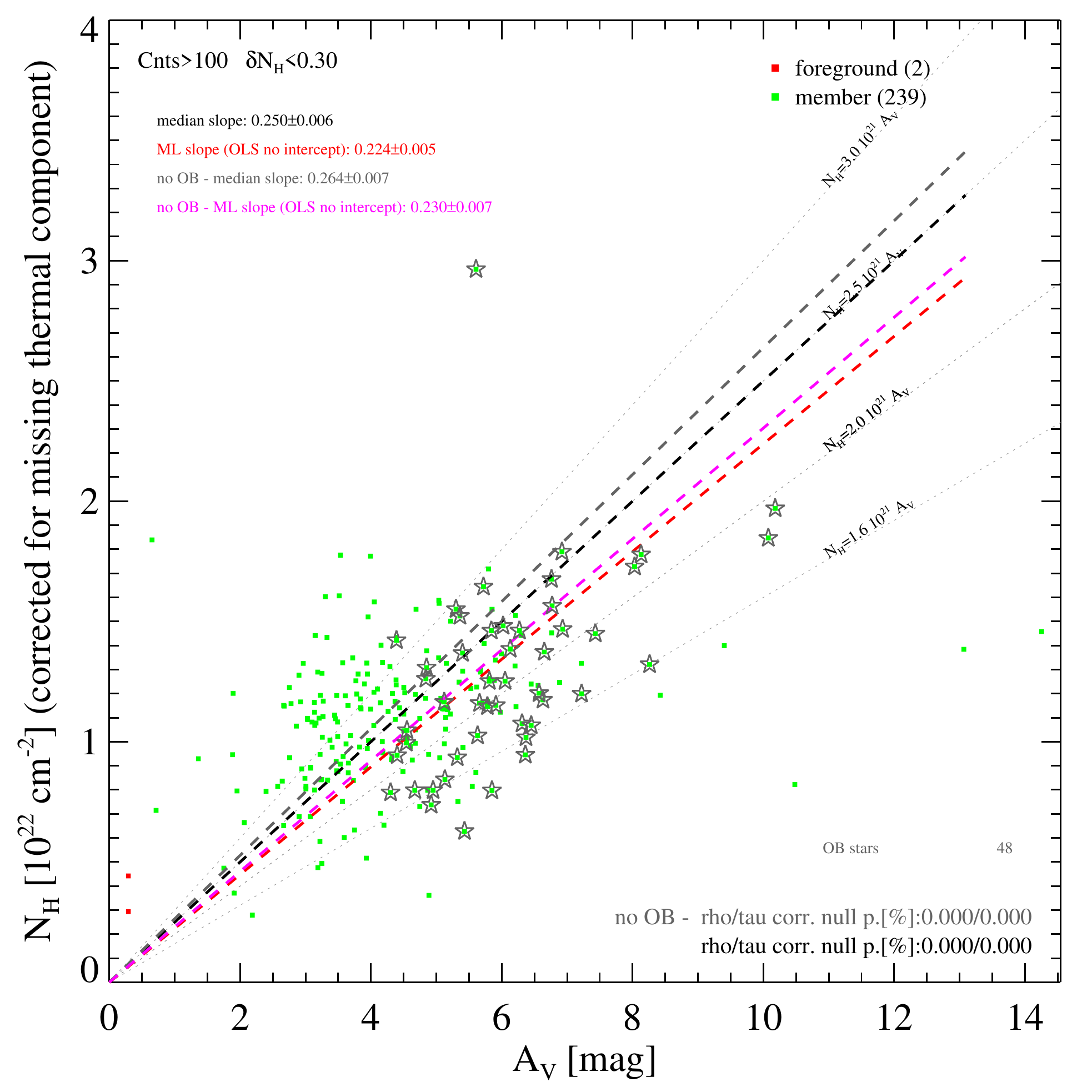}
\caption{[left panel] $N_H$ vs. $A_V$ for Cygnus\,OB2 members with more than 100 X-ray counts and 1$\sigma$ uncertainties on $N_H$ less than 0.3$\times 10^{21}$\,cm$^{-2}$/mag. OB stars are marked by star symbols. Results of Spearman's ($\rho$) and Kendalls's ($\tau$) rank correlation tests are indicated in the bottom-right corner for both the full sample of stars and excluding the OB stars. Two estimates for the slope of the correlation are given in the upper-left corner, again both for the full sample, and excluding the OB stars. These are also illustrated by thick dashed lines. [right panel] same as the left panel but using ``bias''-corrected $N_H$ values.
\label{fig:NHAV}}
\end{figure*}

Our survey of absorbed and reddened stars in Cygnus~OB2 presents a large sample with which to further investigate the gas-to-dust ratio in or toward star-forming regions.
We have used the $N_H$ values from our spectral fits and the $A_V$ values derived from the optical photometry \citep{gua15a} for a new estimate. Both are representative of the total extinction toward the star, including contributions from the ISM toward the Cygnus~OB2 complex, the molecular cloud, and the circumstellar environment (e.g. disks and envelopes). The circumstellar  contribution is, however, unlikely to dominate for Cygnus~OBs members, since stars with and without evidence for circumstellar disks share similarly large extinctions. The molecular cloud is, moreover, also unlikely to dominate the total extinction, since previous studies point toward a large contribution from the Great Cygnus Rift, in the foreground \citep{sal09,got12}. We discuss results obtained adopting $A_V$ values derived by \citet{gua15a} using both the \citet{fuk96} and \citet{fit07} optical extinction laws. Adopting the \citet{fit07} extinction law, the left-hand panel in Fig.\,\ref{fig:NHAV} shows the $N_H$ vs. $A_V$ scatter plot for a sample of 239 Cygnus\,OB2 members with more than 100 detected net X-ray photons and for which the $1\sigma$  uncertainty on $N_H$ was less than 0.30$\times$10$^{22}$\,cm$^{-2}$. Statistical tests indicate that the two quantities are correlated with high confidence. We estimate our best-fit $N_H/A_V$ ratio by both taking the median ratio of all points and performing an ordinary least squares linear fit (with fixed intercept). The results in Figure~\ref{fig:NHAV} are $N_H/A_V=2.0$ and $1.8 \times 10^{21}$\,cm$^{-2}$/mag, respectively for the median ratio and the linear fit, which are compatible with galactic values. Adopting the larger $A_V$ values resulting from the extinction law of \citet{fuk96}, however, the two estimates become, respectively, 1.6 and 1.5$\times 10^{21}$\,cm$^{-2}$/mag, consistent with \citet{vuo03}. Note that Fig.\,\ref{fig:NHAV} also shows the results of the same analysis restricted to low-mass stars, i.e. excluding the 48 OB stars marked by star symbols. These latter stars might in principle behave differently, e.g. because of a different location in the cloud, their clearing of ambient material, and/or the different approach used to estimate $A_V$. The differences in the $N_H/A_V$ slopes with respect to the full sample, however, are not significant.

As already discussed (\S\,\ref{sect:disc_plasma}) and demonstrated in Appendix\,\ref{sect:ONCsim}, however, the isothermal models used to fit the Cygnus~OB2 X-ray spectra, which are statistically adequate owing to their high extinction and poor statistics, introduce significant biases. Using the ONC stars as templates, we simulate the effect of higher extinction and larger distance on the spectral characterization of the sources. We conclude that our $N_H$ values are systematically underestimated by an amount that depends on $N_H$, and which is $\sim 2.0\times 10^{21}$\,cm$^{-2}$, or 20\%, at the typical  extinction of Cygnus\,OB2. We can correct our $N_H$ values for this bias by adding the $N_H$-dependent offsets shown by the black filled symbols in Fig.\,\ref{fig:COUP_sim_Nple1NHFx} (bottom-right panel). The right-hand panel of Fig.\,\ref{fig:NHAV} shows the result of this correction for the case of the $A_V$ values derived from the \citet{fit07} extinction law. The $N_H/A_V$ ratios in this case become  2.5--2.2\,cm$^{-2}$/mag. Even when adopting the \citet{fuk96} optical extinction law, the $N_H/A_V$ ratio increases to 2.0--1.9\,cm$^{-2}$/mag, becoming compatible with the Galactic values.  

We conclude that the stars in Cygnus~OB2 do not support the low value of $N_H/A_V$ found in $\rho$~Ophiuchi by \citet{vuo03} and in the ONC by \citet{has16}.  If those results are correct, then it may point to a variation of the gas-to-dust ratio in different star forming regions and the ONC and $\rho$~Ophiuchi being more dust-rich than Cygnus~OB2. This is perhaps not unreasonable given the very different and milder radiation environments of those regions compared with the intense UV--X-ray and particle radiation fields of Cygnus~OB2 that might be expected to cause more dust destruction \citep[e.g.][]{jon04}. If, however, the extinction toward Cygnus~OB2 is dominated by the Great Cygnus Rift, the $N_H/A_V$ ratio toward the Cygnus~OB2 stars would provide a confirmation of the spatial invariance of the galactic gas-to-dust ratio \citep[e.g.][]{zhu17}.

\section{Summary}
\label{sect:summary_conclusions}

We have derived fluxes and X-ray spectral parameters for the X-ray
sources detected in our {\em Chandra} Cygnus\,OB2 survey, paying 
particular attention to the young members of the region. Most X-ray
sources were detected with few photons and have poorly defined X-ray
spectra. Care was taken to minimize statistical and systematic
uncertainties. For stars with $>$20 counts we fitted isothermal models
with three free spectral parameters; kT, $N_H$, and unabsorbed flux.
However, in order to limit statistical uncertainties for the fainter
Cygnus\,OB2 members, we often restricted the range of allowed
$N_H$ and kT values to the ranges defined by the brighter members. The chemical abundances of the emitting plasma were fixed at the values estimated by \citet{mag07} for similarly young stars in the ONC star forming region. 

A different approach was adopted for X-ray sources detected with $<$20
net counts: 20\% quantiles of the detected photon energy
distributions were demonstrated to be effective indicators of $N_H$ and
of the conversion factor between detected photon-flux and unabsorbed
energy flux. Quantile vs. conversion-factor relations were defined from
the X-ray bright low-mass members with well defined spectral models and
then applied to the fainter sources. This effectively assumes that these fainter sources have X-ray spectra and, in particular, absorbing hydrogen columns, upon which the conversion factors depend most critically, that are similar to those of the brighter sources.

These results of our X-ray spectral fitting will be more thoroughly used in a forthcoming investigation of the general state of magnetic activity in Cygnus~OB2 members.  Here, we have additionally investigated the gas-to-dust ratio toward the Cygnus~OB2 stars by estimating the ratio of X-ray to optical extinction, i.e. the $N_H/A_V$ ratio. We find that, once systematic biases are taken into account, the $N_H/A_V$ ratio toward our Cygnus~OB2 stars is compatible with typical Galactic values, and significantly larger than the value estimated toward other lower mass star forming regions, such as $\rho$ Ophiuchi \citep{vuo03} and the ONC \citep{has16}. 

Finally, we have attempted to verify our assumptions on the plasma abundances through ensemble analyses of the spectra of Cygnus~OB2 members. We conclude that, overall, the abundances of the Cygnus~OB2 stars are compatible with those derived for the slightly younger but similarly active ONC members. In both cases the abundance patterns indicate that an inverse FIP effect is at work, consistent with previous results for very active main sequence stars.


\acknowledgments

MGG and NJW were supported by Chandra Grant GO0-11040X during the course of this work. MGG and EF acknowledge financial contributions from the grant PRIN-INAF 2012 (P.I. E. Flaccomio) and from the ASI-INAF agreement n.2017-14-H.O. NJW acknowledges a Royal Astronomical Society Research Fellowship. 
MM, JJD, and VK were supported by NASA contract NAS8-03060 to the {\it Chandra X-ray Center} 
and thank the director, B. Wilkes, and the CXC science team for continuing support and advice. 
JFAC is a researcher of CONICET and acknowledges their support. 

\software{XSPEC (v12.9; Arnaud 1996), ACIS Extract (Broos et al. 2010)}

\appendix

\section{The ONC at the distance and extinction of Cygnus\,OB2}
\label{sect:ONCsim}

The comparison of X-ray luminosities and spectral parameters of our
Cyg\,OB2 stars with those of young stars in other star forming regions
is complicated by their different distances and extinctions, as well as
by the different observing strategies adopted. The
effect of the relatively large extinction suffered by the Cyg\,OB2
sources ($\sim1\times10^{22}$\,cm$^{-2}$) is particularly relevant, as
it deprives source spectra of their low energy photons, possibly
affecting the reconstruction of intrinsic spectra and the derivation of
plasma temperatures, extinction, and intrinsic X-ray fluxes.

In order to facilitate the above comparison, we here describe Monte Carlo
simulations principally meant to illustrate what the {\em Chandra} observations
described in this paper would have seen had the Cyg\,OB2 population been
replaced by the young stars in the well studied ONC star forming region.  
We will assume that the intrinsic X-ray spectra of the ONC population
are well described by the emission models obtained from the observation
of the {\em Chandra Orion Ultradeep Project} \citep[COUP,][]{get05}.
Given the exposure time of the COUP observation ($\sim$850\,ksec), the
relatively low extinction of the ONC stars, and their closer distance,
this sample is arguably the best characterized among all star forming
regions.

\begin{figure*}[t]
\epsscale{1.17}
\plottwo{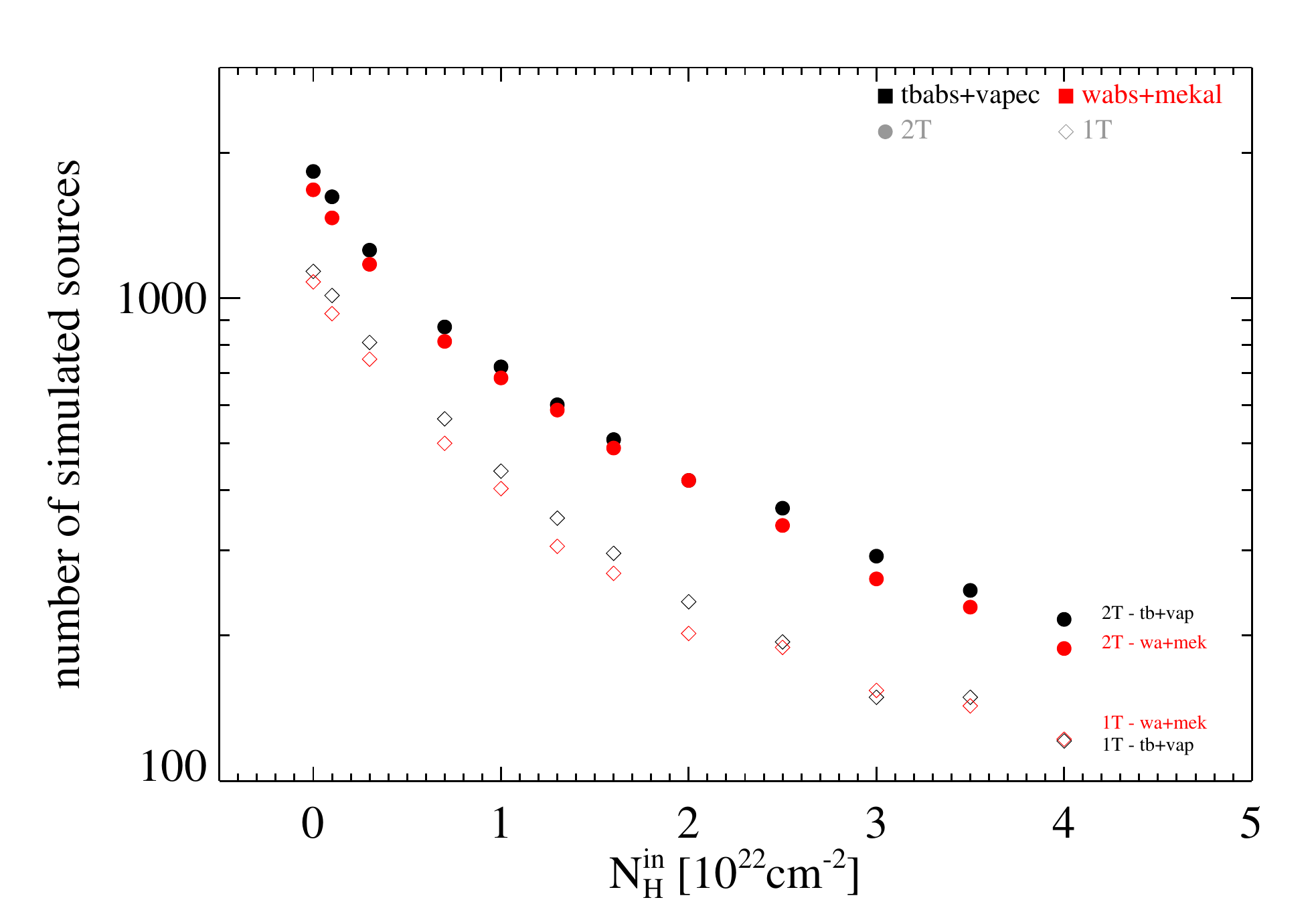}{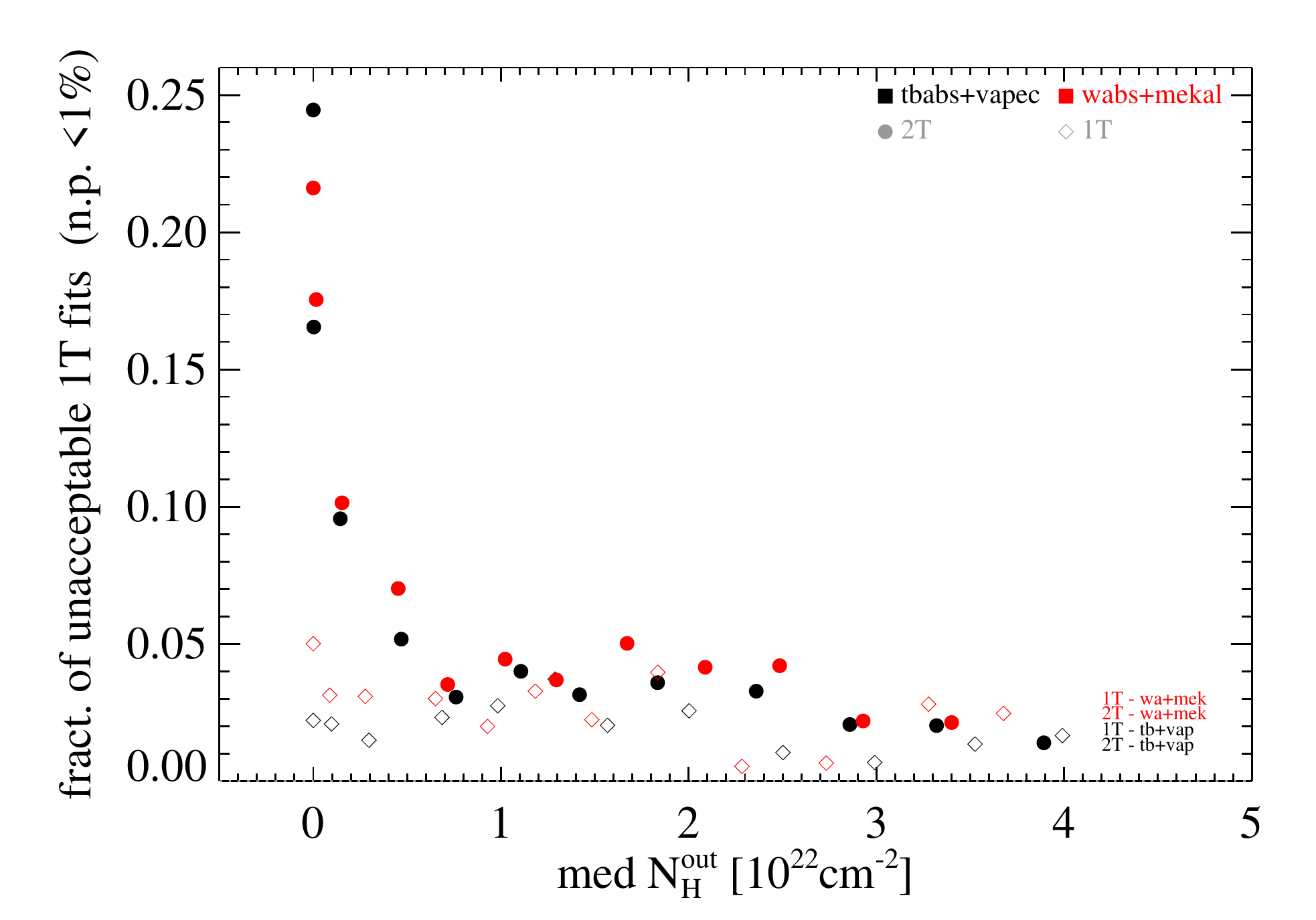}
\plottwo{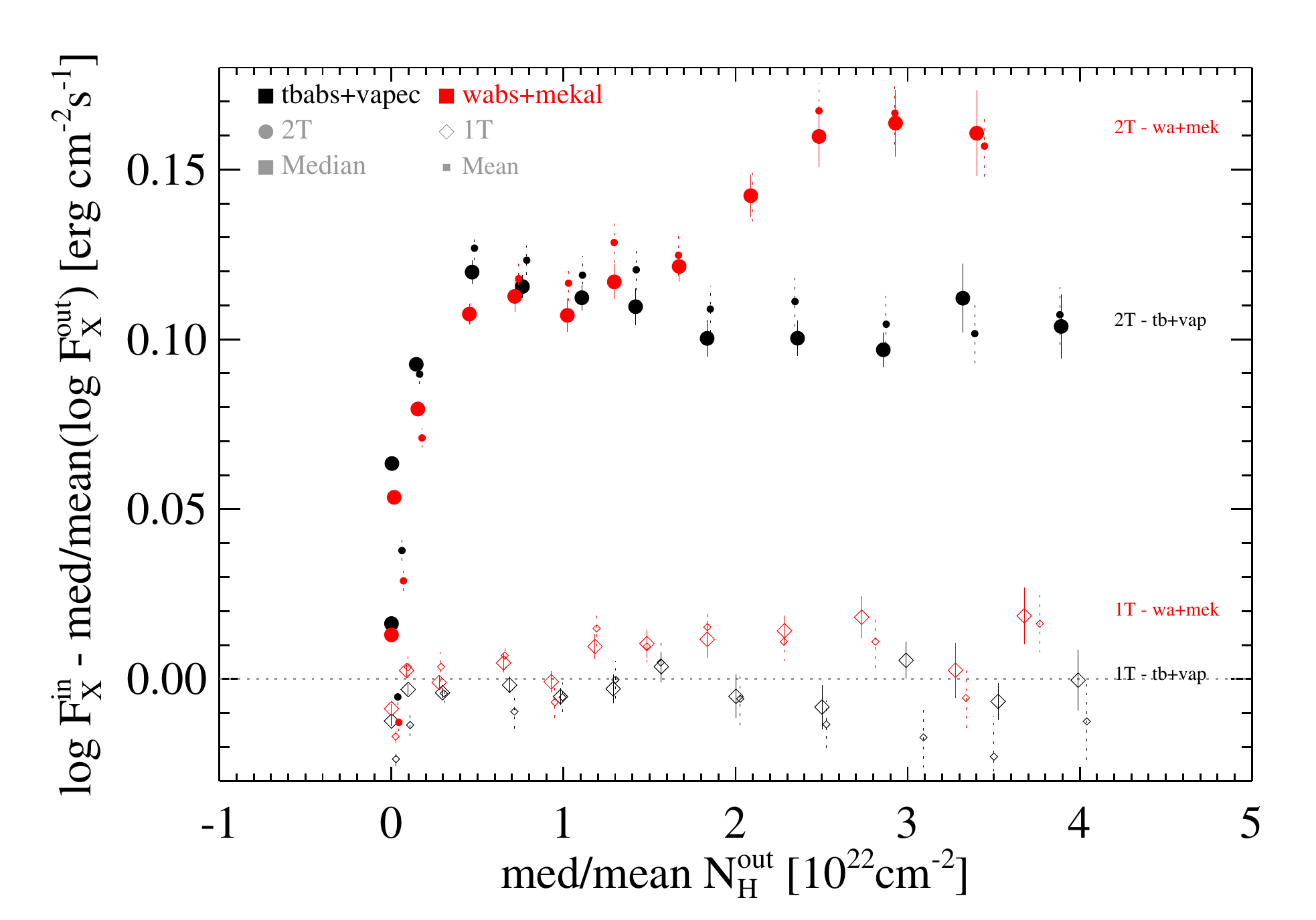}{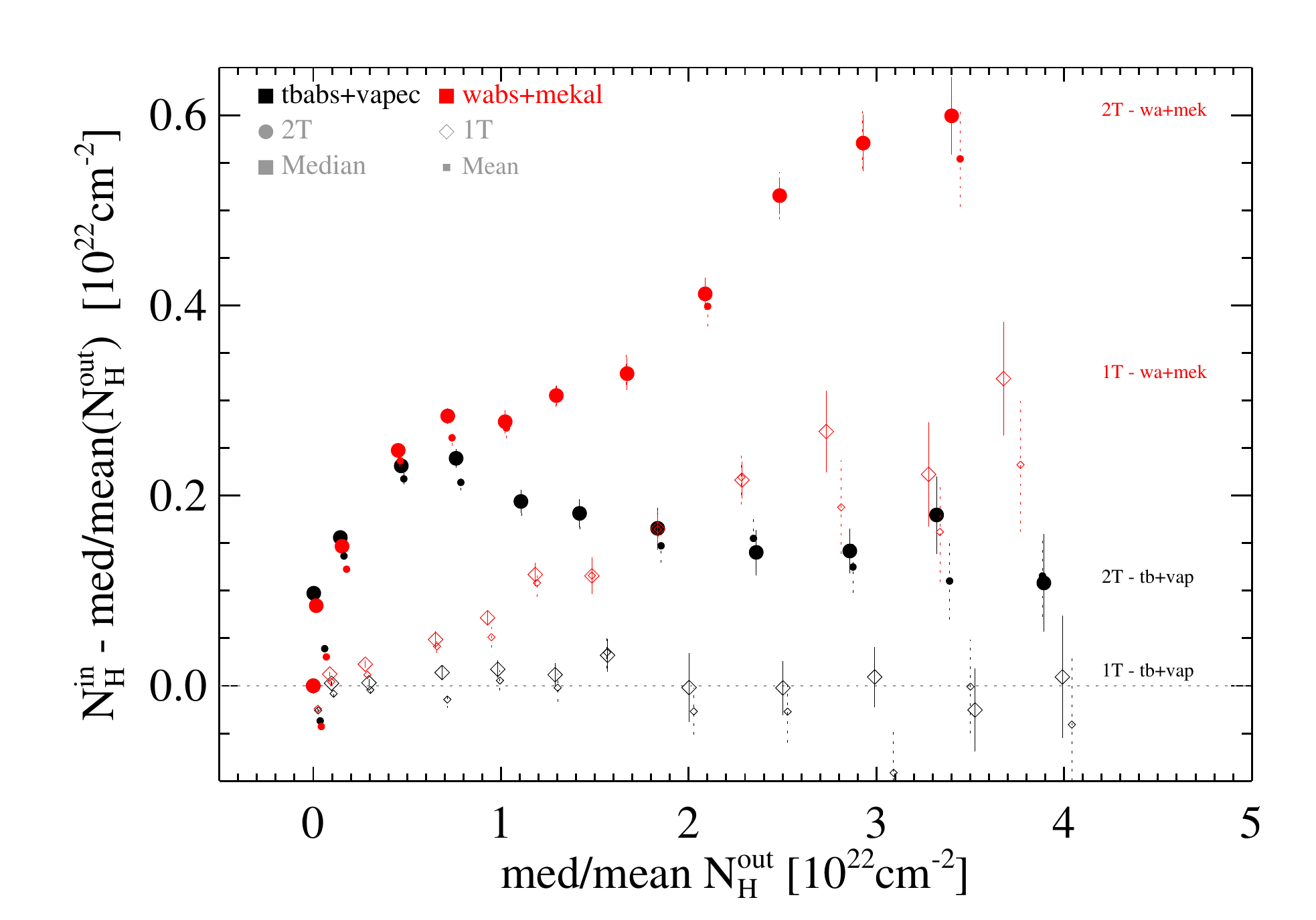}
\caption{Results of simulations to assess the effect of varying X-ray absorption, $N_H$, on the our ability to discriminate between 1T and 2T models, and to retrieve X-ray fluxes and $N_H$ values when fitting X-ray spectra with absorbed isothermal models. The upper-left panel shows the number of simulated sources with more than 100 net counts for each simulation set, i.e. for each value of $N_H^{in}$. Each plotted  set is labeled close to its right-most point and a legend is also provided in the upper-right corner. Figures for two-temperature and isothermal input models are plotted separately, as filled circles and empty diamonds, respectively. Black symbols refer to simulations in which the 1T fitted model is consistent with the input spectrum (1 or 2T), i.e. the absorption and emission models are, respectively, {\sc tbabs} and {\sc vapec} with \citet{mag07} abundances. The red symbols instead refer to the case in which the input spectra are described by the older {\sc wabs} and {\sc mekal} models (with {\sc angr} abundances), while the fitted model is as above. 
Next, in the upper-right panel, we show the fraction of ``unacceptable'' spectral fits as a function of median $N_H^{out}$, again separately for 1T and 2T simulated input spectra and for the two different fit models. Unacceptable fits are defined as those with null probability, as given by the {\em Xspec} {\sc goodness} command, of less than 1\%. Note that the fraction of unacceptable spectral fits, i.e. those for which a second thermal component might improve the fit, is always low for isothermal input models (as expected), as well as for 2T input models for which the $N_H^{out}$ is $\gtrsim 0.7\times10^{22}$\,cm$^{-2}$.
The lower-left panel shows the median (and mean) difference between the reconstructed unabsorbed flux and the flux of the input model as a function of the median (and mean) best fit $N_H$. Large and small symbols, both with error bars, indicate whether median or mean values are considered for each simulation set. The lower-right panel analogously shows, as a function of median/mean best-fit $N_H$, the difference between models and mean/median best-fit $N_H$.  
\label{fig:COUP_sim_Nple1NHFx}}
\end{figure*}

We adopt the following strategy: first, we assign each of the $\sim$1500
COUP sources with an emission model to the position of a random source
in our Cygnus\,OB2 {\em Chandra} catalog and adopt the response function,
effective area, exposure time, and background spectrum of the Cygnus\,OB2
source. We then produce simulated source spectra by adopting the COUP 1-
or 2-temperatures emission models from \citet{get05}\footnote{The
absorption and emission models used in XSPEC were {\sc mekal}, and {\sc
wabs}, respectively, and abundances were set to 0.3 solar, according to
the {\sc angr} tabulation.} but changing $N_H$ to a new value, e.g. to the median
Cyg\,OB2 value ($1.0\times10^{22}$\,cm$^{-2}$) and reducing the
normalization(s) of the emission component(s) by the distance dilution
factor. 
Finally, we fit the spectra with absorbed isothermal models,
using the exact same strategy, models, and abundances we adopted for our
Cyg\,OB2 sources. This process is repeated ten times for each COUP
source, each time adopting the characteristics  (position, exposure
time,  response, background) of a new random Cygnus\,OB2 source. We thus
simulate $\sim$15000 X-ray sources, the great majority with the
X-ray characteristics of $\sim$1\,Myr old stars and spanning the whole
mass range from $\sim$0.1 to $\sim$38\,M$_\odot$. Source spectra were simulated using either one or two thermal components, following the original description of their COUP spectra. In the following, we will mostly focus of on the subset of these spectra with two thermal components, under the reasonable hypothesis that these are the most representative of underlying ``multi-temperature'' spectra. The 1T COUP spectra will, however, serve as a useful cross check, as well as a means to isolate the effect on the best-fit parameters of varying the absorption and emission models.
The simulations were repeated for 12 different $N_H^{out}$ values of the simulated populations, from zero to 4$\times 10^{22}$cm$^{-2}$. For comparison with our Cygnus\,OB2 results, we will focus on the cases with $N_H^{out}\sim 1.0\times 10^{22}$cm$^{-2}$, and consider only the simulated sources with (background-subtracted) simulated spectra containing more than
100 counts.
The resulting samples vary in size depending on the simulation set, as shown in the upper-left panel of Fig.\,\ref{fig:COUP_sim_Nple1NHFx}: for $N_H^{out}\sim 1.0\times 10^{22}cm^{-2}$ the sample with 2T input models and $>$100 simulated counts comprises $\sim$600 simulated sources, which may be compared to the $\sim$400 Cygnus\,OB2 sources observed with $>$100 counts.

For our first experiment we adopt the same emission and absorption
models (vapec and tbabs), and the same coronal abundances \citep{mag07}, for both the simulated spectra and for the fit models. In this way we mean to
test the effects of the increased distance and absorption on the choice of
best-fit model (i.e. 1T vs. 2T) and resulting parameters. The main results of these simulations are shown by black symbols in Fig.\,\ref{fig:COUP_sim_Nple1NHFx}. First we note that, as shown in the upper-right panel of the figure, isothermal fit models generally provide statistically adequate fits to the diluted and absorbed 2T spectra, at least for $N_H^{out}\gtrsim 0.7\times10^{22}$\,cm$^{-2}$. Also, as shown in the two lower panels, we quantify the resulting biases in best-fit $N_H$ values and unabsorbed X-ray fluxes, both of which turn out to be significantly underestimated when fitting 2T spectra with isothermal models. 

For $N_H^{out}=1.0\times10^{22}$\,cm$^{-2}$, i.e. about the median value we obtain for our Cygnus\,OB2 sources, we see that $N_H$ values are actually systematically underestimated by $\sim0.21\times10^{22}$\,cm$^{-2}$ and unabsorbed fluxes by 0.11-0.12\,dex, depending on whether we consider mean or median values.  

Finally, for a more direct comparison with the COUP results, we then repeated the
Monte Carlo simulations adopting the same exact input models provided by
\citet{get05} for the ONC sources, i.e. adopting the {\sc wabs} and {\sc mekal}
absorption and emission models, respectively, and 30\%-solar abundances
according to the \citet{and89} tabulation. These models are most representative of the actual observed COUP spectra. The results, shown by red symbols in Fig.\,\ref{fig:COUP_sim_Nple1NHFx}, were somewhat similar to the previous
set of simulations, at least for $N_H^{out}\sim 1.0\times 10^{22}$\,cm$^{-2}$. We can isolate the effect of changing the absorption and emission model prescriptions by looking at isothermal spectra. We see that the $N_H$ values originally reported by \citet{get05} based on the {\sc wabs$\times$mekal} models with 0.3 times the solar abundances according to \citet{and89}, are systematically lower, by 7-8\%, with respect to those obtained with {\sc tbabs$\times$vapec} and \citet{mag07} abundances.

When looking at best-fit temperatures (not shown), the single temperature fits of two-component spectra yield temperatures that are 0.1-0.2\,keV cooler than the hot component of the input spectrum. At $N_H^{out}\sim 1.0\times 10^{22}$\,cm$^{-2}$, i.e. for the stars of Cygnus\,OB2, the median kT is 0.1\,keV cooler than the hot component and $\sim0.4$\,keV hotter than $kT_{av}$, the average, emission measure-weighted temperature of the input spectra. Since the median $kT_{av}$ of ONC sources with $\log L_X>$30.5\,erg/s (approximatively the lower end for Cygnus~OB2 sources with $>$100 counts) is $\sim$2.4\,keV and the median $kT$ for our Cygnus\,OB2 low mass members with $>$100 counts is $\sim$3.3\,keV, our results seems to indicate that the X-ray emitting plasma  of low-mass stars in the Cygnus~OB2 region might be slightly hotter than in in ONC stars. However, because of the $L_X$-$kT$ correlation (see \S\,\ref{sect:disc_plasma}) and since we are comparing flux limited samples, the difference might be explained by a difference in the composition of the two samples (e.g. due to uncertainties in the distance to Cygnus~OB2). Moreover, a greater incidence of hot flaring emission in the shorter Cygnus~OB2 observations might also explain the small difference in median plasma temperatures. Finally, we stress that the biases we have quantified here should be taken into account when comparing X-ray fluxes and $N_H$ values between regions subject to low and high absorption, such as, respectively, the ONC and Cygnus\,OB2.

\section{Biases from abundance assumptions}
\label{sect:abund_biases}

\begin{figure}[!t!]
\epsscale{1.17}
\plotone{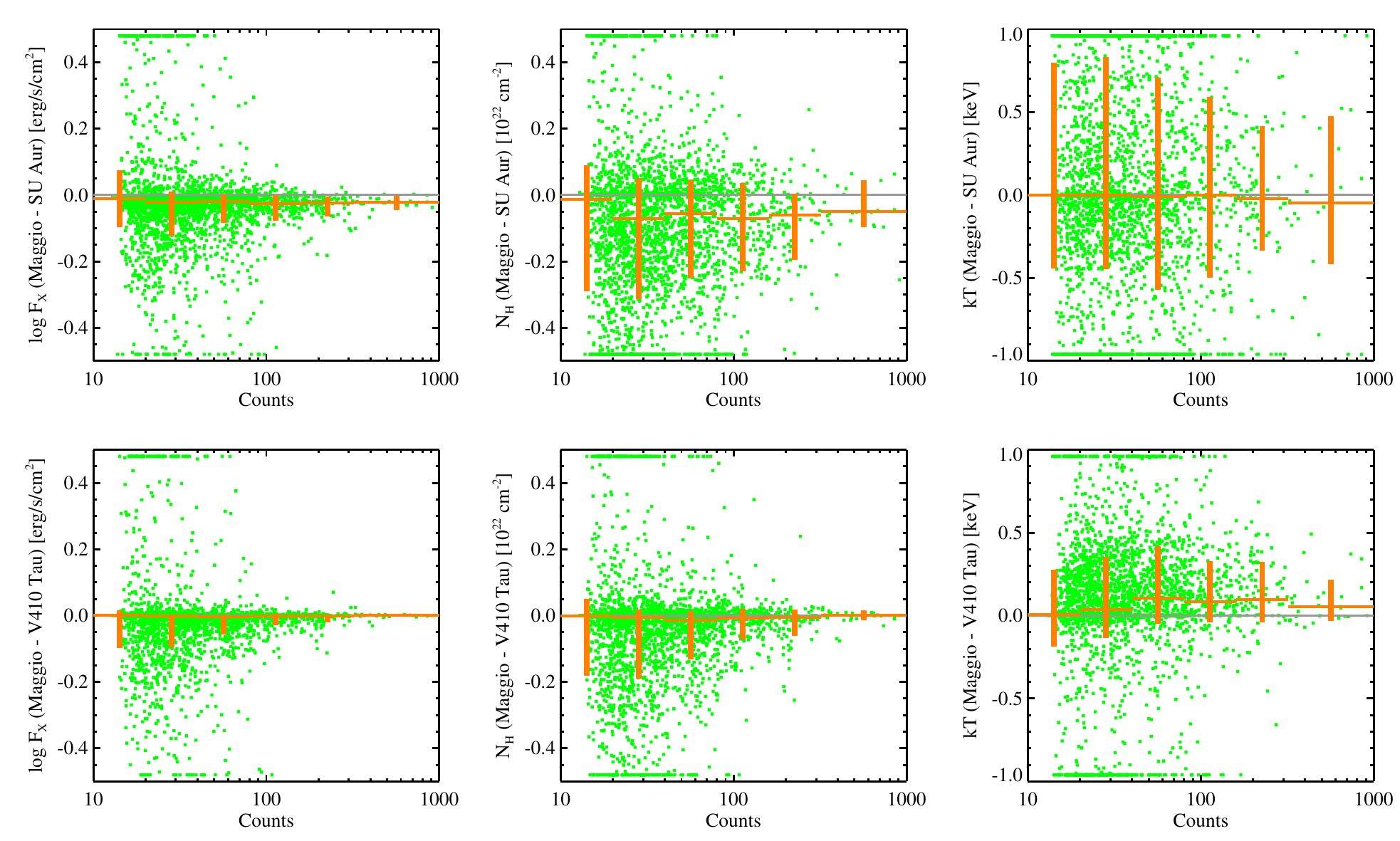}
\caption{Difference between spectral parameters obtained assuming \citet{mag07} abundances and two significantly different abundance sets derived by \citet{tel07a} for SU\,Aur and V410\,Tau (top and bottom rows, respectively). Each panel shows, from left to right, the difference in absorption-corrected flux, $N_H$, and $kT$, as function of net detected counts. Green dots refer to individual Cygnus\,OB2 members (with extreme values plotted close to the top and bottom axes), while orange crosses indicate the median and 1$\sigma$ dispersion of the y-axis values in several count ranges. 
\label{fig:abund_biases}}
\end{figure}

We assess here the effects of different choices of abundance sets for the results of our spectral fitting process. We repeat our spectral analysis (1T fits) using, in addition to our default abundance set \citep{mag07}, those published for SU\,Aur and V410\,Tau in Table\,5 of \citet{tel07a}. Figure\,\ref{fig:abund_biases} summarizes the comparison among the three abundance sets. The upper and lower rows refer to a different pair of abundance sets, Maggio et al. vs. SU Aur and Maggio et al. vs. V410 Tau, respectively. Each column refers to a different fit parameter, from left to right: unabsorbed flux, $N_H$, and kT. Within each panel we plot, for our X-ray sources classified as Cygnus\,OB2 members, the difference between the values of the parameter obtained with the two abundance sets as a function of source net counts. The plots are similar to Fig.\,\ref{fig:NH_kT_counts} in that the orange crosses indicate, in several observed counts ranges, the median and 1$\sigma$ dispersion of the y-axis values. 

We conclude that different choices of abundances make a limited difference for the fit parameters, especially when considering median values. Individual sources, especially low signal-to-noise ratio ones, will end-up  with somewhat different parameters depending on the adopted abundance sets. We verified, however, that these differences are smaller than the 1$\sigma$ uncertainties on our best-fit parameters (obtained with the Maggio et al. abundances) in $>$94\% and $>$98\% of the cases, when adopting the SU Aur and the V410 Tau abundances, respectively.

\bibliographystyle{apj} 
\bibliography{bibtex.bib}

\begin{thebibliography}{}
\expandafter\ifx\csname natexlab\endcsname\relax\def\natexlab#1{#1}\fi

\bibitem[{{Albacete-Colombo} {et~al.}(2016){Albacete-Colombo}, {Flaccomio},
  {Drake}, {Wright}, {Guarcello}, \& {Kashyap}}]{alb16}
{Albacete-Colombo}, J.~F., {Flaccomio}, E., {Drake}, J.~J., {et~al.} 2016,
  ArXiv e-prints, arXiv:1603.08372

\bibitem[{{Albacete Colombo} {et~al.}(2007){Albacete Colombo}, {Flaccomio},
  {Micela}, {Sciortino}, \& {Damiani}}]{alb07}
{Albacete Colombo}, J.~F., {Flaccomio}, E., {Micela}, G., {Sciortino}, S., \&
  {Damiani}, F. 2007, \aap, 464, 211

\bibitem[{{Anders} \& {Grevesse}(1989)}]{and89}
{Anders}, E., \& {Grevesse}, N. 1989, \gca, 53, 197

\bibitem[{{Arnaud}(1996)}]{arn96}
{Arnaud}, K.~A. 1996, in Astronomical Society of the Pacific Conference Series,
  Vol. 101, Astronomical Data Analysis Software and Systems V, ed. G.~H.
  {Jacoby} \& J.~{Barnes}, 17--+

\bibitem[{{Brinkman} {et~al.}(2001){Brinkman}, {Behar}, {G{\"u}del}, {Audard},
  {den Boggende}, {Branduardi-Raymont}, {Cottam}, {Erd}, {den Herder},
  {Jansen}, {Kaastra}, {Kahn}, {Mewe}, {Paerels}, {Peterson}, {Rasmussen},
  {Sakelliou}, \& {de Vries}}]{bri01}
{Brinkman}, A.~C., {Behar}, E., {G{\"u}del}, M., {et~al.} 2001, \aap, 365, L324

\bibitem[{{Broos} {et~al.}(2010){Broos}, {Townsley}, {Feigelson}, {Getman},
  {Bauer}, \& {Garmire}}]{bro10}
{Broos}, P.~S., {Townsley}, L.~K., {Feigelson}, E.~D., {et~al.} 2010, \apj,
  714, 1582

\bibitem[{{Cash}(1979)}]{cas79}
{Cash}, W. 1979, \apj, 228, 939

\bibitem[{{Comer{\'o}n} {et~al.}(2002){Comer{\'o}n}, {Pasquali}, {Rodighiero},
  {Stanishev}, {De Filippis}, {L{\'o}pez Mart{\'{\i}}}, {G{\'a}lvez Ortiz},
  {Stankov}, \& {Gredel}}]{com02a}
{Comer{\'o}n}, F., {Pasquali}, A., {Rodighiero}, G., {et~al.} 2002, \aap, 389,
  874

\bibitem[{{Da Rio} {et~al.}(2010){Da Rio}, {Robberto}, {Soderblom}, {Panagia},
  {Hillenbrand}, {Palla}, \& {Stassun}}]{dar10}
{Da Rio}, N., {Robberto}, M., {Soderblom}, D.~R., {et~al.} 2010, \apj, 722,
  1092

\bibitem[{{Drake}(2002)}]{dra02}
{Drake}, J.~J. 2002, in Astronomical Society of the Pacific Conference Series,
  Vol. 277, Stellar Coronae in the Chandra and XMM-NEWTON Era, ed. F.~{Favata}
  \& J.~J. {Drake}, 75

\bibitem[{{Drake}(2003)}]{dra03a}
{Drake}, J.~J. 2003, Advances in Space Research, 32, 945

\bibitem[{{Drake} {et~al.}(2001){Drake}, {Brickhouse}, {Kashyap}, {Laming},
  {Huenemoerder}, {Smith}, \& {Wargelin}}]{dra01}
{Drake}, J.~J., {Brickhouse}, N.~S., {Kashyap}, V., {et~al.} 2001, \apjl, 548,
  L81

\bibitem[{{Drake} {et~al.}(1996){Drake}, {Laming}, \& {Widing}}]{dra96}
{Drake}, J.~J., {Laming}, J.~M., \& {Widing}, K.~G. 1996, in IAU Colloq. 152:
  Astrophysics in the Extreme Ultraviolet, ed. S.~{Bowyer} \& R.~F. {Malina},
  97

\bibitem[{{Drew} {et~al.}(2008){Drew}, {Greimel}, {Irwin}, \& {Sale}}]{dre08a}
{Drew}, J.~E., {Greimel}, R., {Irwin}, M.~J., \& {Sale}, S.~E. 2008, \mnras,
  386, 1761

\bibitem[{{Feigelson} {et~al.}(2005){Feigelson}, {Getman}, {Townsley},
  {Garmire}, {Preibisch}, {Grosso}, {Montmerle}, {Muench}, \&
  {McCaughrean}}]{fei05}
{Feigelson}, E.~D., {Getman}, K., {Townsley}, L., {et~al.} 2005, \apjs, 160,
  379

\bibitem[{{Feldmeier} {et~al.}(1997){Feldmeier}, {Puls}, \&
  {Pauldrach}}]{fel97}
{Feldmeier}, A., {Puls}, J., \& {Pauldrach}, A.~W.~A. 1997, \aap, 322, 878

\bibitem[{{Fitzpatrick} \& {Massa}(2007)}]{fit07}
{Fitzpatrick}, E.~L., \& {Massa}, D. 2007, \apj, 663, 320

\bibitem[{{Fruscione} {et~al.}(2006){Fruscione}, {McDowell}, {Allen},
  {Brickhouse}, {Burke}, {Davis}, {Durham}, {Elvis}, {Galle}, {Harris},
  {Huenemoerder}, {Houck}, {Ishibashi}, {Karovska}, {Nicastro}, {Noble},
  {Nowak}, {Primini}, {Siemiginowska}, {Smith}, \& {Wise}}]{fru06}
{Fruscione}, A., {McDowell}, J.~C., {Allen}, G.~E., {et~al.} 2006, in
  \procspie, Vol. 6270, Society of Photo-Optical Instrumentation Engineers
  (SPIE) Conference Series, 62701V

\bibitem[{{Fukugita} {et~al.}(1996){Fukugita}, {Ichikawa}, {Gunn}, {Doi},
  {Shimasaku}, \& {Schneider}}]{fuk96}
{Fukugita}, M., {Ichikawa}, T., {Gunn}, J.~E., {et~al.} 1996, \aj, 111, 1748

\bibitem[{{Getman} {et~al.}(2005){Getman}, {Flaccomio}, {Broos}, {Grosso},
  {Tsujimoto}, {Townsley}, {Garmire}, {Kastner}, {Li}, {Harnden}, {Wolk},
  {Murray}, {Lada}, {Muench}, {McCaughrean}, {Meeus}, {Damiani}, {Micela},
  {Sciortino}, {Bally}, {Hillenbrand}, {Herbst}, {Preibisch}, \&
  {Feigelson}}]{get05}
{Getman}, K.~V., {Flaccomio}, E., {Broos}, P.~S., {et~al.} 2005, \apjs, 160,
  319

\bibitem[{{Getman} {et~al.}(2011){Getman}, {Broos}, {Feigelson}, {Townsley},
  {Povich}, {Garmire}, {Montmerle}, {Yonekura}, \& {Fukui}}]{get11}
{Getman}, K.~V., {Broos}, P.~S., {Feigelson}, E.~D., {et~al.} 2011, \apjs, 194,
  3

\bibitem[{{Gottschalk} {et~al.}(2012){Gottschalk}, {Kothes}, {Matthews},
  {Landecker}, \& {Dent}}]{got12}
{Gottschalk}, M., {Kothes}, R., {Matthews}, H.~E., {Landecker}, T.~L., \&
  {Dent}, W.~R.~F. 2012, \aap, 541, A79

\bibitem[{{Guarcello} {et~al.}(2015){Guarcello}, {Drake}, {Wright}, {Naylor},
  {Flaccomio}, {Kashyap}, \& {Garcia-Alvarez}}]{gua15a}
{Guarcello}, M.~G., {Drake}, J.~J., {Wright}, N.~J., {et~al.} 2015, ArXiv
  e-prints, arXiv:1501.03761

\bibitem[{{Guarcello} {et~al.}(2013){Guarcello}, {Drake}, {Wright}, {Drew},
  {Gutermuth}, {Hora}, {Naylor}, {Aldcroft}, {Fruscione},
  {Garc{\'{\i}}a-Alvarez}, {Kashyap}, \& {King}}]{gua13a}
---. 2013, \apj, 773, 135

\bibitem[{{Guarcello} {et~al.}(2016){Guarcello}, {Drake}, {Wright},
  {Albacete-Colombo}, {Clarke}, {Ercolano}, {Flaccomio}, {Kashyap}, {Micela},
  {Naylor}, {Schneider}, {Sciortino}, \& {Vink}}]{gua16a}
---. 2016, ArXiv e-prints, arXiv:1605.01773

\bibitem[{{G{\"u}del} {et~al.}(2007){G{\"u}del}, {Skinner}, {Mel'Nikov},
  {Audard}, {Telleschi}, \& {Briggs}}]{gud07}
{G{\"u}del}, M., {Skinner}, S.~L., {Mel'Nikov}, S.~Y., {et~al.} 2007, \aap,
  468, 529

\bibitem[{{Hanson}(2003)}]{han03a}
{Hanson}, M.~M. 2003, \apj, 597, 957

\bibitem[{{Hasenberger} {et~al.}(2016){Hasenberger}, {Forbrich}, {Alves},
  {Wolk}, {Meingast}, {Getman}, \& {Pillitteri}}]{has16}
{Hasenberger}, B., {Forbrich}, J., {Alves}, J., {et~al.} 2016, \aap, 593, A7

\bibitem[{{Hillenbrand}(1997)}]{hil97}
{Hillenbrand}, L.~A. 1997, \aj, 113, 1733

\bibitem[{{Hong} {et~al.}(2004){Hong}, {Schlegel}, \& {Grindlay}}]{hon04}
{Hong}, J., {Schlegel}, E.~M., \& {Grindlay}, J.~E. 2004, \apj, 614, 508

\bibitem[{{Johnstone} \& {G{\"u}del}(2015)}]{joh15a}
{Johnstone}, C.~P., \& {G{\"u}del}, M. 2015, ArXiv e-prints, arXiv:1505.00643

\bibitem[{{Jones}(2004)}]{jon04}
{Jones}, A.~P. 2004, in Astronomical Society of the Pacific Conference Series,
  Vol. 309, Astrophysics of Dust, ed. A.~N. {Witt}, G.~C. {Clayton}, \& B.~T.
  {Draine}, 347

\bibitem[{{Kashyap} {et~al.}(2018){Kashyap}, {Guarcello}, {Wright}, {Drake}, \&
  {et al.}}]{kas18}
{Kashyap}, V., {Guarcello}, M.~G., {Wright}, N.~J., {Drake}, J.~J., \& {et al.}
  2018, ApJ (under review)

\bibitem[{{Kiminki} \& {Kobulnicky}(2012)}]{kim12a}
{Kiminki}, D.~C., \& {Kobulnicky}, H.~A. 2012, \apj, 751, 4

\bibitem[{{Kiminki} {et~al.}(2007){Kiminki}, {Kobulnicky}, {Kinemuchi},
  {Irwin}, {Fryer}, {Berrington}, {Uzpen}, {Monson}, {Pierce}, \&
  {Woosley}}]{kim07a}
{Kiminki}, D.~C., {Kobulnicky}, H.~A., {Kinemuchi}, K., {et~al.} 2007, \apj,
  664, 1102

\bibitem[{{Laming}(2009)}]{lam09}
{Laming}, J.~M. 2009, \apj, 695, 954

\bibitem[{{Laming}(2015)}]{lam15}
---. 2015, Living Reviews in Solar Physics, 12, 2

\bibitem[{{Laming} {et~al.}(1995){Laming}, {Drake}, \& {Widing}}]{lam95}
{Laming}, J.~M., {Drake}, J.~J., \& {Widing}, K.~G. 1995, \apj, 443, 416

\bibitem[{{Maggio} {et~al.}(2007){Maggio}, {Flaccomio}, {Favata}, {Micela},
  {Sciortino}, {Feigelson}, \& {Getman}}]{mag07}
{Maggio}, A., {Flaccomio}, E., {Favata}, F., {et~al.} 2007, \apj, 660, 1462

\bibitem[{{Massey} \& {Thompson}(1991)}]{mas91a}
{Massey}, P., \& {Thompson}, A.~B. 1991, \aj, 101, 1408

\bibitem[{{Peres} {et~al.}(2004){Peres}, {Orlando}, \& {Reale}}]{per04}
{Peres}, G., {Orlando}, S., \& {Reale}, F. 2004, \apj, 612, 472

\bibitem[{{Predehl} \& {Schmitt}(1995)}]{pre95}
{Predehl}, P., \& {Schmitt}, J.~H.~M.~M. 1995, \aap, 293, 889

\bibitem[{{Preibisch} {et~al.}(2005{\natexlab{a}}){Preibisch}, {Kim}, {Favata},
  {Feigelson}, {Flaccomio}, {Getman}, {Micela}, {Sciortino}, {Stassun},
  {Stelzer}, \& {Zinnecker}}]{pre05}
{Preibisch}, T., {Kim}, Y.-C., {Favata}, F., {et~al.} 2005{\natexlab{a}},
  \apjs, 160, 401

\bibitem[{{Preibisch} {et~al.}(2005{\natexlab{b}}){Preibisch}, {McCaughrean},
  {Grosso}, {Feigelson}, {Flaccomio}, {Getman}, {Hillenbrand}, {Meeus},
  {Micela}, {Sciortino}, \& {Stelzer}}]{pre05a}
{Preibisch}, T., {McCaughrean}, M.~J., {Grosso}, N., {et~al.}
  2005{\natexlab{b}}, \apjs, 160, 582

\bibitem[{{Rauw} {et~al.}(2015){Rauw}, {Naz{\'e}}, {Wright}, {Drake},
  {Guarcello}, {Prinja}, {Peck}, {Albacete Colombo}, {Herrero}, {Kobulnicky},
  {Sciortino}, \& {Vink}}]{rau15}
{Rauw}, G., {Naz{\'e}}, Y., {Wright}, N.~J., {et~al.} 2015, \apjs, 221, 1

\bibitem[{{Ryter}(1996)}]{ryt96}
{Ryter}, C.~E. 1996, \apss, 236, 285

\bibitem[{{Sale} {et~al.}(2009){Sale}, {Drew}, {Unruh}, {Irwin}, {Knigge},
  {Phillipps}, {Zijlstra}, {G{\"a}nsicke}, {Greimel}, {Groot}, {Mampaso},
  {Morris}, {Napiwotzki}, {Steeghs}, \& {Walton}}]{sal09}
{Sale}, S.~E., {Drew}, J.~E., {Unruh}, Y.~C., {et~al.} 2009, \mnras, 392, 497

\bibitem[{{Schneider} {et~al.}(2006){Schneider}, {Bontemps}, {Simon}, {Jakob},
  {Motte}, {Miller}, {Kramer}, \& {Stutzki}}]{sch06}
{Schneider}, N., {Bontemps}, S., {Simon}, R., {et~al.} 2006, \aap, 458, 855

\bibitem[{{Schulte}(1956)}]{sch56a}
{Schulte}, D.~H. 1956, \apj, 124, 530

\bibitem[{{Skinner} \& {G{\"u}del}(2017)}]{ski17}
{Skinner}, S.~L., \& {G{\"u}del}, M. 2017, \apj, 839, 45

\bibitem[{{Telleschi} {et~al.}(2007{\natexlab{a}}){Telleschi}, {G{\"u}del},
  {Briggs}, {Audard}, \& {Palla}}]{tel07}
{Telleschi}, A., {G{\"u}del}, M., {Briggs}, K.~R., {Audard}, M., \& {Palla}, F.
  2007{\natexlab{a}}, \aap, 468, 425

\bibitem[{{Telleschi} {et~al.}(2007{\natexlab{b}}){Telleschi}, {G{\"u}del},
  {Briggs}, {Audard}, \& {Scelsi}}]{tel07a}
{Telleschi}, A., {G{\"u}del}, M., {Briggs}, K.~R., {Audard}, M., \& {Scelsi},
  L. 2007{\natexlab{b}}, \aap, 468, 443

\bibitem[{{Vink} {et~al.}(2008){Vink}, {Drew}, {Steeghs}, {Wright}, {Martin},
  {G{\"a}nsicke}, {Greimel}, \& {Drake}}]{vin08a}
{Vink}, J.~S., {Drew}, J.~E., {Steeghs}, D., {et~al.} 2008, \mnras, 387, 308

\bibitem[{{Vuong} {et~al.}(2003){Vuong}, {Montmerle}, {Grosso}, {Feigelson},
  {Verstraete}, \& {Ozawa}}]{vuo03}
{Vuong}, M.~H., {Montmerle}, T., {Grosso}, N., {et~al.} 2003, \aap, 408, 581

\bibitem[{{Wood} {et~al.}(2018){Wood}, {Laming}, {Warren}, \&
  {Poppenhaeger}}]{woo18}
{Wood}, B.~E., {Laming}, J.~M., {Warren}, H.~P., \& {Poppenhaeger}, K. 2018,
  ArXiv e-prints, arXiv:1806.05111

\bibitem[{{Wright} {et~al.}(2016){Wright}, {Bouy}, {Drew}, {Sarro}, {Bertin},
  {Cuillandre}, \& {Barrado}}]{wri16a}
{Wright}, N.~J., {Bouy}, H., {Drew}, J.~E., {et~al.} 2016, \mnras, 460, 2593

\bibitem[{{Wright} \& {Drake}(2009)}]{wri09}
{Wright}, N.~J., \& {Drake}, J.~J. 2009, \apjs, 184, 84

\bibitem[{{Wright} {et~al.}(2012){Wright}, {Drake}, {Drew}, {Guarcello},
  {Gutermuth}, {Hora}, \& {Kraemer}}]{wri12a}
{Wright}, N.~J., {Drake}, J.~J., {Drew}, J.~E., {et~al.} 2012, \apjl, 746, L21

\bibitem[{{Wright} {et~al.}(2010){Wright}, {Drake}, {Drew}, \& {Vink}}]{wri10a}
{Wright}, N.~J., {Drake}, J.~J., {Drew}, J.~E., \& {Vink}, J.~S. 2010, \apj,
  713, 871

\bibitem[{{Wright} {et~al.}(2014{\natexlab{a}}){Wright}, {Drake}, {Guarcello},
  {Aldcroft}, {Kashyap}, {Damiani}, {DePasquale}, \& {Fruscione}}]{wri14a}
{Wright}, N.~J., {Drake}, J.~J., {Guarcello}, M.~G., {et~al.}
  2014{\natexlab{a}}, ArXiv e-prints, arXiv:1408.6579

\bibitem[{{Wright} {et~al.}(2015{\natexlab{a}}){Wright}, {Drake}, {Guarcello},
  {Kashyap}, \& {Zezas}}]{wri15b}
{Wright}, N.~J., {Drake}, J.~J., {Guarcello}, M.~G., {Kashyap}, V.~L., \&
  {Zezas}, A. 2015{\natexlab{a}}, ArXiv e-prints, arXiv:1511.03943

\bibitem[{{Wright} {et~al.}(2015{\natexlab{b}}){Wright}, {Drew}, \&
  {Mohr-Smith}}]{wri15a}
{Wright}, N.~J., {Drew}, J.~E., \& {Mohr-Smith}, M. 2015{\natexlab{b}}, \mnras,
  449, 741

\bibitem[{{Wright} {et~al.}(2014{\natexlab{b}}){Wright}, {Parker}, {Goodwin},
  \& {Drake}}]{wri14b}
{Wright}, N.~J., {Parker}, R.~J., {Goodwin}, S.~P., \& {Drake}, J.~J.
  2014{\natexlab{b}}, \mnras, 438, 639

\bibitem[{{Zhu} {et~al.}(2017){Zhu}, {Tian}, {Li}, \& {Zhang}}]{zhu17}
{Zhu}, H., {Tian}, W., {Li}, A., \& {Zhang}, M. 2017, \mnras, 471, 3494

\end{thebibliography}



\end{document}